\documentclass[aip,jcp,reprint]{revtex4-1}
\usepackage{graphicx}
\usepackage{dcolumn}
\usepackage{bm}
\usepackage{mathrsfs}
\usepackage{amssymb}
\usepackage[fleqn]{amsmath}
\usepackage{color}
\newcommand{\Liou}{\mathcal{L_{}}}
\DeclareMathOperator{\Tr} { Tr}
\allowdisplaybreaks

\begin{document}

\title{Molecular theory of Langevin dynamics for active self-diffusiophoretic
colloids}

\author{Bryan Robertson}
\email{bryan.robertson@mail.utoronto.ca}
\affiliation{ Chemical Physics Theory Group, Department of Chemistry, University of Toronto, Toronto, Ontario M5S 3H6, Canada}

\author{Jeremy Schofield}
\email{jmschofi@chem.utoronto.ca}
\affiliation{ Chemical Physics Theory Group, Department of Chemistry, University of Toronto, Toronto, Ontario M5S 3H6, Canada}

\author{Pierre Gaspard}
\email{gaspard@ulb.ac.be}
\affiliation{ Center for Nonlinear Phenomena and Complex Systems, Universit{\'e} Libre de Bruxelles (U.L.B.), Code Postal 231, Campus Plaine, B-1050 Brussels, Belgium}

\author{Raymond Kapral}
\email{rkapral@chem.utoronto.ca}
\affiliation{ Chemical Physics Theory Group, Department of Chemistry, University of Toronto, Toronto, Ontario M5S 3H6, Canada}

\date{\today}


\begin{abstract}
Active colloidal particles that are propelled by a self-diffusiophoretic mechanism are often described by Langevin equations that are either postulated on physical grounds or derived using the methods of fluctuating hydrodynamics. While these descriptions are appropriate for colloids of micrometric and larger size, they will break down for very small active particles. A fully microscopic derivation of Langevin equations for self-diffusiophoretic particles powered by chemical reactions catalyzed asymmetrically by the colloid is given in this paper. The derivation provides microscopic expressions for the translational and rotational friction tensors, as well as reaction rate coefficients appearing in the Langevin equations. The diffusiophoretic force and torque are expressed in terms of nonequilibrium averages of fluid fields that satisfy generalized transport equations. The results provide a description of active motion on small scales where descriptions in terms of coarse grained continuum fluid equations combined with boundary conditions that account for the presence of the colloid may not be appropriate.
\end{abstract}

\maketitle

\section{Introduction}
Active matter systems can take many forms, ranging from those whose active agents are microorganisms or synthetic colloids to active materials and gels, among many others.~\cite{R10,V12,A13,EWG15,FM18,ZS16} Since active matter is not at equilibrium its properties  often differ markedly from its equilibrium analogs, and this fact has prompted investigations that explore the mechanisms by which such systems function and their possible applications.

Here we consider active colloidal particles that are self-propelled through a diffusiophoretic mechanism where chemical reactions, maintained out of equilibrium, take place on a catalyst that is asymmetrically distributed on the colloid and produce concentration gradients in reactants and products.~\cite{DD74,A86,A89,ALP82,GLA05,K13,CRRK14,PUD16,BDLRVV16,S19} Interactions of the colloid with chemical species under these nonequilibrium conditions give rise to fluid flows in the vicinity of the colloid as a consequence of momentum conservation, leading to propulsion of the active particle.

Active colloidal particles with micrometer sizes are frequently considered in experiments~\cite{W13,WDAMS13,SSK15,WDS2016,AWQF16} so that continuum descriptions of the fluid in which they move are adequate; however, on this length scale thermal fluctuations cannot be neglected. As a result stochastic descriptions, usually in the form of Langevin equations, are used to describe the motions of these particles. In its simplest form the Langevin equation that describes the evolution of the velocity $\bm{V}$ of an active colloidal particle with mass $M$ propelled by a self-diffusiophoretic mechanism is written as~\cite{GK19}
\begin{equation}
M \frac{d}{dt}\bm{V}= \bm{F}_{\rm sd} -\zeta_t \bm{V} +\bm{F}_{\rm fl},
\end{equation}
where $\zeta_t$ is a friction coefficient, $\bm{F}_{\rm fl}$ is a random force and the new ingredient that distinguishes this equation from that for simple equilibrium Brownian motion is $\bm{F}_{\rm sd}$, the diffusiophoretic force. Under most conditions the inertial term on the left can be neglected for micrometric particles in condensed phases and the overdamped limit of this equation is sufficient. The expression for the diffusiophoretic force, or the corresponding diffusiophoretic velocity, $\bm{V}_{\rm sd}=\bm{F}_{\rm sd}/\zeta_t$ in overdamped descriptions, is often simply postulated or derived~\cite{GK18a} from continuum models of the fluid subject to boundary conditions that account for coupling to the colloid.

On smaller nanometer or even {\AA}ngstr{\"o}m scales continuum descriptions will break down since the dimensions of fluid particles may no longer be negligible on the scale of the colloid size. In these cases where the molecular nature of the fluid manifests itself in the vicinity of the colloid it is difficult to describe fluid-colloid interactions through boundary conditions. Active colloids with linear dimensions on the order of a few tens of nanometers have been studied in the laboratory.~\cite{LAMHGF14,APTW14} While motions of these very small active particles are dominated by thermal noise, the characteristics of active motion persist and are observable. In addition, molecular dynamics simulations of very small active dimer colloids with linear dimensions of a few nanometers exhibit features of active motion due to catalytic chemical reactions on part of their surface.~\cite{CK14} Even for these very small particles the local fluid velocity fields, obtained by extensive averaging to remove thermal noise effects, show flow patterns that are characteristic of self-diffusiophoresis. This feature is reminiscent of the fluid velocity fields observed in early molecular dynamics simulations of tagged particle motion that lead to long-time power law decay of velocity correlations.~\cite{AW67,AW70} Such collective solvent motions contribute to the values of diffusion coefficients and form the microscopic basis for Stokes law relating the frictional force on the colloid to the viscosity of the solvent.~\cite{DBM76,CKLM80,SO92} In a similar way the microscopic flow fields seen in the vicinities of tiny active particles point to the presence of coupling to fluid collective modes with hydrodynamic character and the operation of a diffusiophoretic mechanism on molecular scales.

In order to study Brownian motion on very small scales where continuum descriptions break down, a molecular perspective must be adopted, and molecular derivations of Langevin equations for inactive colloidal particles have been carried out.  Perhaps the most complete description is that of Mazur and Oppenheim~\cite{MO70} where the statistical properties of the noise are determined for a massive Brownian particle in an equilibrium bath. Such derivations have been extended to situations where the fluid in which the Brownian particle moves is subjected to constraints that drive it out of equilibrium.~\cite{SO96,ED15}

Similarly, to study active motion on very small scales a molecular description is needed where the particulate nature of the solvent is taken into account and assumptions on the large relative colloid to solute size are relaxed. In this paper we present a molecular derivation of the Langevin equations that describe the translational and rotational dynamics of a rigid active self-diffusiophoretic colloidal particle in a nonequilibrium environment. Since the system must be out of equilibrium for active motion to take place, we make use of a statistical mechanical formulation that accounts for the constraints that maintain the system in a nonequilibrium state.

A Langevin description of the translational and orientational dynamics of a colloidal particle is obtained from the equations of motion for the entire system by projecting out the bath degrees of freedom. Because the bath is in a nonequilibrium state a time-dependent projection operator formalism is required, where the projection operator averages dynamical variables over a nonequilibrium bath density that depends conditionally on the presence of a fixed colloid. The nonequilibrium density is expressed in terms a local equilibrium density containing time-dependent local thermodynamic fields conjugate to microscopic hydrodynamic density fields.  The conjugate fields are defined self-consistently by constraint conditions that require the nonequilibrium averages of the hydrodynamic densities to be given exactly at all points in the system by averages over the local equilibrium density.

In addition, since our description is fully microscopic, we show how to include catalytic reactive dynamics in a way that treats the reactive chemical species at a molecular level. The resulting generalized Langevin equations serve the dual functions of describing active diffusiophoretic dynamics on molecular scales and providing microscopic expressions for the transport properties the enter Langevin descriptions on larger scales.

Section~\ref{sec1:langevin} of the paper specifies the system comprising the colloid and its fluid environment, gives an expression for its Hamiltonian and presents the Liouville equations that govern its evolution. Chemical species are defined in Sec.~\ref{sec:chemical-reactions} in terms of microscopic reaction coordinates and species variables that depend on the internal coordinates of the reactive molecules. The densities and constraints that characterize and determine the nonequilibrium state of the system are presented in Sec.~\ref{sec:nonequilibrium}. The derivation of the generalized Langevin equations using nonequilibrium time-dependent projection operator methods is given in Sec.~\ref{sec:LangevinDerivation}, while in Sec.~\ref{sec:linear-angular} it is shown how these general equations yield the Langevin equations for the linear and angular momenta of the active colloid. The diffusiophoretic force and torque that are responsible for the active motion are further discussed in Sec.~\ref{sec:averageF}, and Sec.~\ref{sec:conclusion} gives the conclusions of the study. Additional details of the calculations are presented in the Appendices.

\section{System and dynamics}\label{sec1:langevin}

The physical system considered here consists of a single rigid colloid of arbitrary mass distribution and total mass $M$ immersed in a multi-component fluid of molecules of mass $m$.~\cite{colloid}  The fluid is composed of reactive molecules dilutely dispersed in a solvent in contact with reservoirs that isothermally feed and remove species from the system at boundaries that are spatially distant from the colloid.  A typical configuration of the physical system consists of $N_R$ molecules of the reactive species $R$ and $N_S$ solvent molecules $S$ with $N_S \gg N_R$. These fluid species are denoted by $\nu \in \{S,R\}$. Each reactive molecule $i$ with total mass $m$ contains $n_a$ chemically bound atoms with masses $\{m_k\;|\; k=1,2,\dots, n_a\}$ and nuclear positions and momenta $\bm{x}_{i}^{n_a}=(\bm{r}_{(1)i},\bm{p}_{(1)i}, \dots, \bm{r}_{(n_a)i},\bm{p}_{(n_a)i}) =(\bm{r}_{i}^{n_a},\bm{p}_{i}^{n_a})$. The coordinates and momenta of the collection of the $N_R$ reactive molecules are denoted by $\bm{x}_m^{N_R}=(\bm{x}^{n_a}_1, \dots, \bm{x}^{n_a}_{N_R})=(\bm{r}_m^{N_R},\bm{p}_m^{N_R})$. While the solvent molecules, also taken to have mass $m$ for simplicity,  can be described in a similar way, their internal degrees of freedom will play no role in this work and only their center-of-mass positions and momenta will be considered, $\bm{x}^{N_S}=(\bm{r}_{N_R+1},\bm{p}_{N_R+1},...,\bm{r}_{N_R+N_S},\bm{p}_{N_R+N_S}) =(\bm{r}^{N_S},\bm{p}^{N_S})$.

\begin{figure}[htbp]
\centering
\resizebox{1.0\columnwidth}{!}{%
\includegraphics{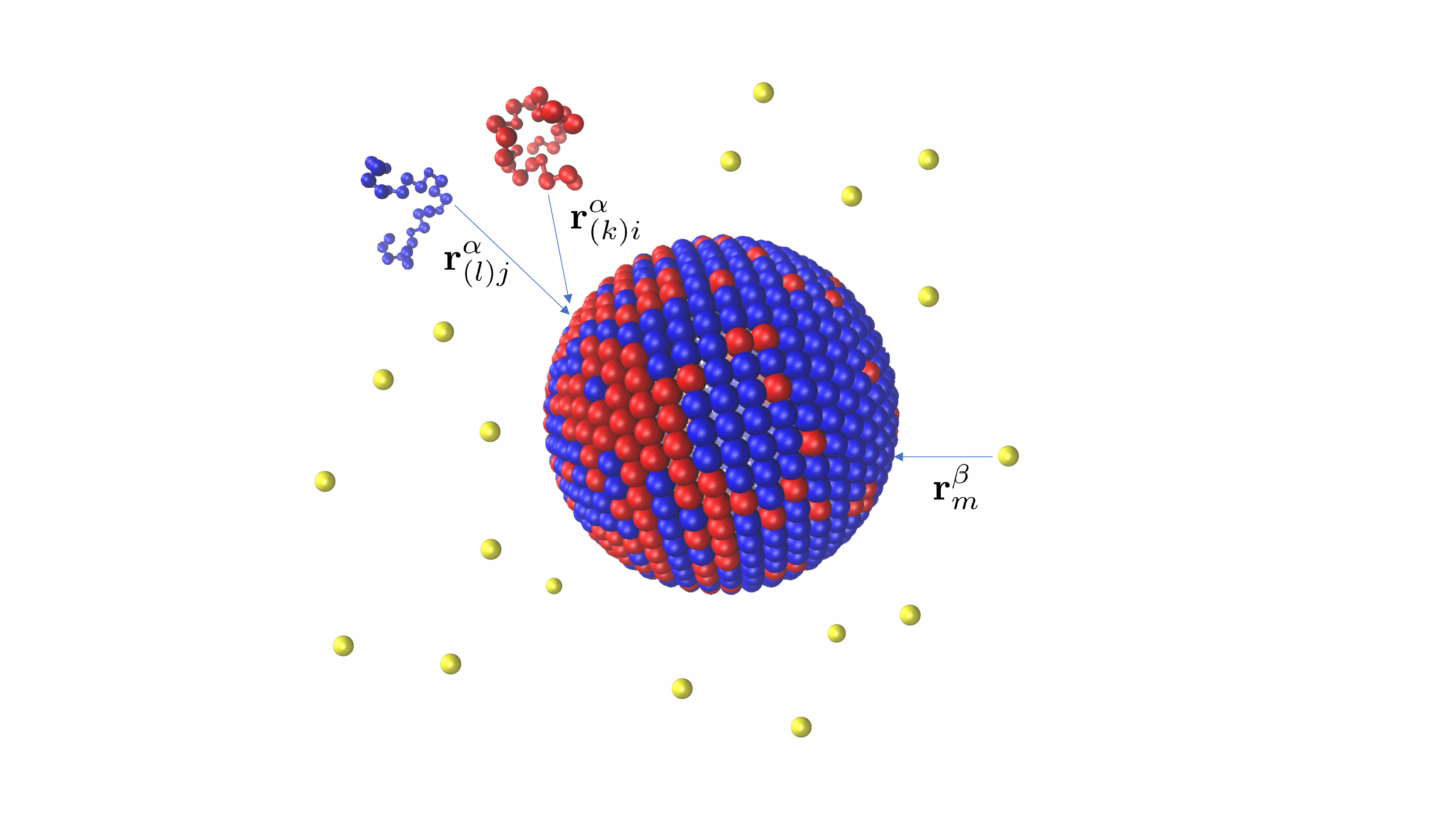}}
\caption{\label{fig:system} Illustration showing the components of the system: Solvent particles are represented by yellow spheres, reactive molecules of type $A$ (red) and $B$ (blue) are composed of $n_a$ atoms, and the colloid possesses both catalytic (red) and noncatalytic (blue) sites.  In this graphic, the interaction sites are configured to represent a spherical colloid comprised of irregularly-distributed catalytic and noncatalytic sites but other geometries and distributions can be considered. }
\end{figure}

The spherical colloid has a total of $n_s$ catalytic $C$ and noncatalytic $N$ sites on its surface. The distribution of these sites on the surface is left arbitrary at this point and may be chosen to describe active colloids with various properties. For instance, if the catalytic sites are confined to one hemisphere the colloid is a Janus particle. Although it is feasible to treat the internal motions of the components of the colloid to allow for energy exchange between the internal degrees of freedom of the colloid and the surrounding fluid molecules, we will assume that the small and rapid internal fluctuations of the positions of the components of the colloid around their equilibrium values are not physically important in an isothermal system.  For the rigid model the positions of the sites relative to the center of the colloid, $\bm{S}^{\alpha} (\bm{R})$, are at a fixed distance from the center of the colloid, and can be written as $\bm{S}^\alpha(\bm{R})\equiv \bm{S}^\alpha - \bm{R} = \bm{A}^T \cdot \tilde{\bm{S}}^\alpha$, where $\bm{A}^T$ is a rotation matrix known as the attitude matrix that converts vectors between body-fixed and laboratory frames of reference, and the $\tilde{\bm{S}}^\alpha$ are constant vectors specifying the location of a site $\alpha$ relative to the center of the colloid in the body-fixed frame.\cite{vanZS07,vanZS08} The rotation matrix $\bm{A}^T$ and its inverse $\bm{A}$ are specified by a set of arbitrary Euler orientational angles $\bm{\theta}$. In general, the active colloid need not be significantly larger than the solvent in which it moves. Figure~\ref{fig:system} shows the type of active colloid, reactive molecules with internal structure and structureless solvent molecules comprising the system under study.

In writing the sums over particles it is convenient to define indicator functions $\Theta_i^{\nu}$ where $\Theta_i^{\nu}=1$ if molecule $i$ is species $\nu$ and $\Theta_i^{\nu}=0$ otherwise. Using this notation to determine whether molecule $i$ is a solvent molecule or a reactive solute, the nuclear Hamiltonian for a system  with $N$ fluid molecules may be written as
\begin{eqnarray}\label{eq:Ham_bath_col}
H&=&\frac{P^2}{2M}+K_{\rm rot} + \sum_{i=1}^N \Theta_i^S\frac{p_{i}^{2}}{2m}+\sum_{i=1}^N \Theta_i^R H_{mi} \nonumber\\
&&+U_{\rm f}(\bm{r}^{N_S},\bm{r}_m^{N_R})+U_{\rm I}(\bm{R},\bm{r}^{N_S},\bm{r}_m^{N_R}).
\end{eqnarray}
This Hamiltonian is the sum of the translational and rotational kinetic energies of the colloid, the kinetic energies of the centers of mass of the $N_S$ solvent molecules and the sum of the reactive molecule Hamiltonians,
\begin{equation}\label{eq:molecularH}
H_{mi}= \sum_{k=1}^{n_a} \Big(\frac{p_{(k)i}^2}{2m_k} + V_m(\bm{r}_i^{n_a})\Big),
\end{equation}
where $V_m(\bm{r}^{n_a}_i)$ is the potential function for the nuclei in chemically-bonded molecule $i$. Interactions among the fluid molecules are given by $U_{\rm f}$, while $U_{\rm I}$ describes the interactions of the fluid particles with the colloid.

In the laboratory frame, the time derivative of the {\it relative} site position vector $\bm{S}^\alpha(\bm{R}) = \bm{A}^T \cdot \tilde{\bm{S}}^\alpha$ is given in terms of the angular velocities $\bm{\omega}$ by $\dot{\bm{S}}^\alpha = \bm{\omega}  \wedge \bm{S}^\alpha (\bm{R})= \dot{\bm{\theta}}^T \cdot \bm{\nabla}_\theta \bm{A}^T  \cdot
\bm{A} \cdot (\bm{S}^\alpha -\bm{R})$, from which one finds that the angular velocities are related to time derivatives of the angles by $\bm{\omega} = \bm{N}^T \cdot \dot{\bm{\theta}}$, where the elements of the matrix $\bm{N}$ are
\begin{equation}
N_{ab} = \frac{1}{2} \epsilon^{bcd} A_{ec}
 \,\nabla_{\theta_a} A_{ed} .
\label{Nmatrixdef}
\end{equation}
Here $\epsilon^{bcd}$ is the Levi-Civita symbol and the Einstein convention of a sum over repeated indices has been used. The rotational kinetic energy of the colloid is~\cite{Goldstein80}
\begin{equation}
K_{\rm rot} = \frac{1}{2} \bm{\omega}^T \cdot \bm{I_m} \cdot \bm{\omega} =
\frac{1}{2} \dot{\bm{\theta}}^T \cdot \bm{M}  \cdot \dot{\bm{\theta}},
\label{eq:rotationalKE}
\end{equation}
where $\bm{I_m}$ is the moment of inertia tensor in the laboratory frame and the matrix $\bm{M} = \bm{N} \cdot \bm{I_m} \cdot \bm{N}^T$.  Defining the generalized momentum $\bm{\Pi}$ conjugate to the angles $\bm{\theta}$ as $\bm{\Pi} = \partial K_{\rm rot} / \partial \dot{\bm{\theta}} = \bm{M} \cdot \dot{\bm{\theta}}$, the total Hamiltonian in Eq.~(\ref{eq:Ham_bath_col}) for the system with colloidal phase space coordinates $\bm{X} = (\bm{R}, \bm{P}, \bm{\theta}, \bm{\Pi})$ can now be written as
\begin{equation}
H= \frac{P^2}{2M} + \frac{1}{2} \bm{\Pi}^T \cdot \bm{M}^{-1} \cdot \bm{\Pi}+ H_0,
\label{Hamiltonian-with-rotation}
\end{equation}
 which defines $H_0$, the bath Hamiltonian in the presence of the fixed colloidal particle. It will play a central role in the development that follows.

The bath Hamiltonian $H_0$ contains the $V_m$, $U_{\rm f}$ and $U_{\rm I}$ potential functions. While the potential function for the chemically-bonded atoms in a molecule, $V_m(\bm{r}^{n_a}_i)$, is generally a many-body potential, we assume that the non-bonded interactions between the atoms in different molecules as well as those between the atoms in  a molecule and the solvent molecules are pair-wise additive. Consequently, we can write
\begin{eqnarray}
U_{\rm f} &=&  \sum_{i=1}^N U_{{\rm f} i}=\sum_{i=1}^N \Big[\frac{1}{2} \sum_{\substack{j=1\\(i \ne j)}}^{N} \Big( \Theta_i^S \Theta_j^S V_{SS}(r_{ij})\nonumber \\
 &&+ 2\Theta_i^S \Theta_j^R \sum_{k=1}^{n_a} V_{Sk}(|\bm{r}_i-\bm{r}_{(k)j}|) \nonumber \\
&&+   \Theta_i^R \Theta_j^R \sum_{k,k'=1}^{n_a} V_{kk'}(|\bm{r}_{(k)i}-\bm{r}_{(k')j}|)\Big)\Big].
\end{eqnarray}

We also assume that the non-bonded interactions between the solvent and atoms in the reactive molecules with the $n_s$ sites on the colloid are pair-wise additive. Then, the $U_{\rm I}$ interaction potential can be written as
\begin{eqnarray}\label{eq:U-interaction}
U_{\rm I} &=& \sum_{i=1}^{N}\Big[ \sum_{\alpha =1}^{n_s}\sum_{b=N}^C  \Theta_\alpha^b  \Big(\Theta_i^S V_{Sb}({r}_{i}^\alpha) +\Theta_i^R \sum_{k=1}^{n_a} V_{kb}({r}_{(k)i}^{\alpha})\Big)\Big],\nonumber \\
&=& \sum_{i=1}^{N} U_{{\rm I}i}=\sum_{\alpha =1}^{n_s} U_{\rm I}^\alpha.
\end{eqnarray}
Here $\bm{r}_{i}^\alpha = \bm{r}_i-{\bm{S}}^\alpha =\bm{r}_{ic}-{\bm{S}}^\alpha (\bm{R})$ and $\bm{r}_{(k)i}^\alpha=\bm{r}_{(k)i}-{\bm{S}}_\alpha =\bm{r}_{(k)ic}-{\bm{S}}_\alpha (\bm{R})$ where $\bm{r}_{ic}=\bm{r}_i-\bm{R}$ and $\bm{r}_{(k)ic}=\bm{r}_{(k)i}-\bm{R}$ are the center-of-mass and atom positions of molecule $i$ relative to the center of mass of the colloid. In the last line of Eq.~(\ref{eq:U-interaction}) we interchanged the sums on fluid particles and colloid sites to define $U_{\rm I}^\alpha$, the interaction potential for the solvent molecules with the site $\alpha$ on the colloid.  Interactions of the fluid molecules with the colloidal sites are taken to be short-ranged and are zero beyond a cut-off distance $\sigma_c$ from the colloid center.

\subsection{Time evolution}

The time evolution of a dynamical variable $B(\bm{x}^{N_S},\bm{x}_m^{N_R},\bm{X})$ is given by the Liouville equation
\begin{equation}\label{eq:liou}
\partial_t B(t)= -\{H,B(t)\} = i\Liou \, B(t),
\end{equation}
where $i\Liou$, the Liouville operator for the evolution of the entire system, is defined in terms of the Poisson bracket of the Hamiltonian and the dynamical variable. It can be written as $i\Liou =i\Liou_c +i\Liou_0$, the sum of the Liouvillian for the colloid, $i\Liou_c$, and the Liouvillian for the bath in the presence of the fixed colloid, $i\Liou{}_0$. The Liouvillian for the colloid is
\begin{equation}
i\Liou{}_c= \frac{\bm{P}}{M} \cdot \bm{\nabla_R} -  \bm{\nabla_R} U_{\rm I} \cdot \bm{\nabla_P} + i{\cal L}_{\rm rot}.
\end{equation}
The rotational part of the Liouville operator ${\cal L}_{\rm rot}$ can be decomposed into an operator for the free rotation of a rigid body and an operator for the orientationally-dependent interactions,
\begin{eqnarray}
i{\mathcal L}_{\rm rot} &=& i{\mathcal L}_{\rm rot,f} - \bm{\nabla}_{\bm{\theta}} U_{\rm I}\cdot \bm{\nabla}_{\bm{\Pi}} \nonumber \\
i{\mathcal L}_{\rm rot,f} &=& \bm{\Pi}^T \cdot \bm{M}^{-1} \cdot \bm{\nabla_\theta} - \bm{\nabla_\theta} K_{\rm rot} \cdot \bm{\nabla_{\Pi}}.
\label{rotationalLiouville}
\end{eqnarray}
The Liouville operator $i{\mathcal L}_{\rm rot,f}$ for the free rotation of a rigid body has the property that $i{\mathcal L}_{\rm rot,f} \, \bm{L} = 0$, where $\bm{L} = \bm{I_m} \cdot \bm{\omega} = \bm{N}^{-1} \cdot \bm{\Pi}$ is the angular momentum of the colloid.~\cite{vanZS07} The torque on the colloid, $\bm{T}$,  is given by the time derivative of the angular momentum vector,
\begin{equation}
 \bm{T}=\dot{\bm{L}} = - \bm{\nabla_\theta} U_{\rm I} \cdot \bm{\nabla_\Pi} \left( \bm{N}^{-1} \cdot \bm{\Pi} \right) = - \bm{N}^{-1} \cdot \bm{\nabla_\theta} U_{\rm I}.
\label{angularMomentumderiv}
\end{equation}
The force on the colloid, $\bm{F}_c$, is given by the time derivative of the momentum,
\begin{equation}
\bm{F}_c=\dot{\bm{P}}=-\bm{\nabla_R}U_{\rm I}.
\end{equation}

The Liouvillian for the bath in the presence of the colloid is,
\begin{eqnarray}
i\Liou_0 &=& \sum_{i=1}^N \Theta_i^S \Big(\frac{\bm{p}_i}{m} \cdot \bm{\nabla}_{\bm{r}_i} -\bm{\nabla}_{\bm{r}_i} (U_{\rm f}+U_{\rm I}) \cdot \bm{\nabla}_{\bm{p}_i} \Big)\nonumber \\
&&+\sum_{i=1}^N \Theta_i^R \sum_{k=1}^{n_a}\Big( \frac{\bm{p}_{(k)i}}{m_k} \cdot \bm{\nabla}_{\bm{r}_{(k)i}} \nonumber \\
&& \quad -\bm{\nabla}_{\bm{r}_{(k)i}} (U_m+U_{\rm f}+U_{\rm I}) \cdot \bm{\nabla}_{\bm{p}_{(k)i}} \Big),
\end{eqnarray}
where $U_m=\sum_{i=1}^N \Theta_i^R V_m(\bm{r}_i^{n_a})$.

\section{Chemical reactions and species densities}\label{sec:chemical-reactions}

The motions of active colloids that operate by a self-diffusiophoretic mechanism are powered by catalytic chemical reactions on their surfaces using fuel supplied by chemical species in their environments. The uncatalyzed reactions among reactive molecules that take place in the fluid far from the colloid are assumed to be controlled by high free energy barriers so that reactive events are very infrequent and are often neglected on the time scales on which the colloidal dynamics occurs. However, when these species interact with the catalytic portions of the colloid the free energy barriers that control the reaction rates are significantly reduced, facilitating more rapid interconversion among reactants and products, thus allowing the diffusiophoretic mechanism to operate. In experiments, the catalysts can vary widely, ranging from metals to enzymes, and the corresponding reactive fuel species vary from frequently-used hydrogen peroxide to the substrates specific to given enzymes.~\cite{W13,WDAMS13,SSK15,WDS2016,AWQF16}

The description of reactive dynamics from a microscopic perspective entails the derivation of macroscopic rate laws from the microscopic equations of motion for chemical species densities specified at a molecular level.~\cite{KCM98} Since the chemical species change their identities under the dynamics, they are metastable molecular states. For condensed phase reactions it is sufficient to use configuration space criteria to define them, and their specification may involve the use of one or more reaction coordinates that depend on the reaction mechanism.~\cite{CCHK89,CKV05} While the details are system dependent and their implementations may vary in difficulty, the basic aspects of the formulation presented here can be generalized to accommodate a variety of reaction mechanisms; e.g., those involving bimolecular reactions or various surface reactions. Here we illustrate the application of the formalism with a simple chemical reaction, $A \rightleftharpoons B$, where interactions with the colloid allow fuel $A$ and product $B$ species to interconvert.

\begin{figure}[htbp]
\centering
\resizebox{1.0\columnwidth}{!}{%
\includegraphics{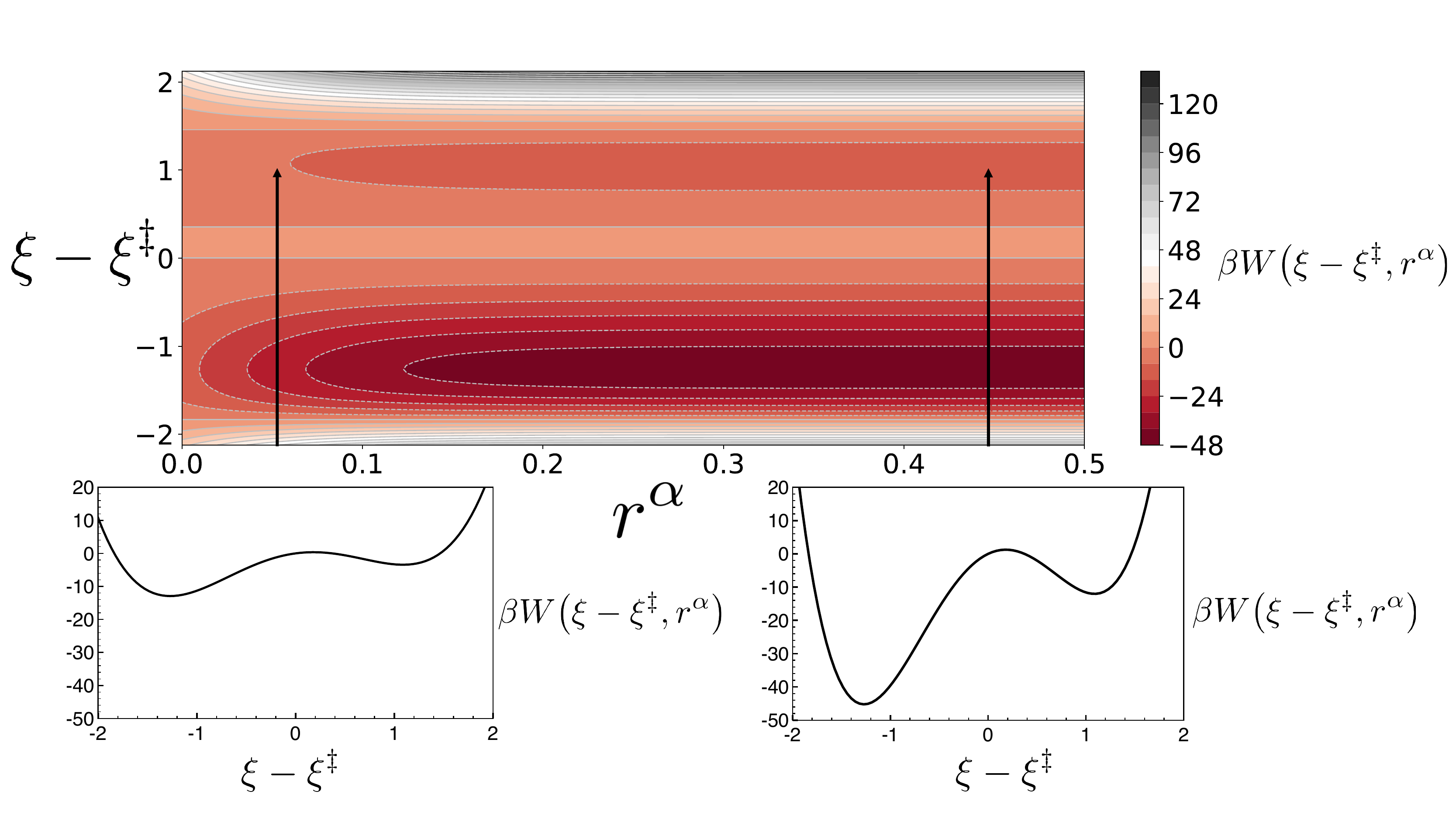}}
\caption{\label{fig:W} The upper part of the figure plots $W(\xi,r^\alpha)$ as color-coded function of $\xi$ and $r^\alpha$. It shows the potential wells corresponding to the metastable $A$ and $B$ species separated by a free energy barrier. The lower panels show how $W(\xi,r^\alpha)$ varies with $\xi$ at two chosen values of $r^\alpha$: the lower right panel is for an $r^\alpha$ value where the reactive molecule is far from the colloid and $W(\xi,r^\alpha)$ has double-well structure with deep wells separated by a high barrier, while the lower left panel is for an $r^\alpha$  value where the molecule interacts with colloid and the barrier separating the two metastable states is low and reaction is much more likely that in the bulk fluid. In this schematic figure the numbers on the axis labels are simply guides to illustrate the changes in the well depths and barrier heights.}
\end{figure}

Specifically, the reactive molecules are assumed to exist in two long-lived metastable states characterized by two distinct sets of nuclear configurations corresponding to the $A$ and $B$ chemical species. The metastable $A$ and $B$ species can be specified by introducing a scalar reaction coordinate, $\xi_i(\bm{r}_{i}^{n_a})$, that is used to define a hypersurface $\xi_i(\bm{r}_{i}^{n_a})=\xi^\ddagger$ in the configuration space of the molecule that separates regions where the metastable chemical species lie. In order to describe the change in the reaction dynamics when the reactive molecules interact with the colloid, it is useful to introduce a second scalar reaction coordinate that is the distance of the center of mass of the reactive molecule from an active site on the colloid, $r^\alpha_{i}(\bm{r}_{i}^{n_a})= |\bm{r}_i-{\bm{S}}^\alpha| =|\bm{r}_{ic}-{\bm{S}}^\alpha (\bm{R})|$ as defined earlier but now the center of mass of a reactive molecule is given by $\bm{r}_i=\sum_{k=1}^{n_a} (m_k/m) \bm{r}_{(k)i}$. The free energy along the vectorial reaction coordinate $( \xi_i(\bm{r}_{i}^{n_a}),r^\alpha_{i}(\bm{r}_{i}^{n_a}))$ can be defined as $W(\xi,r^\alpha) =-\beta^{-1} \ln (P(\xi,r^\alpha)/P_u)$, with the probability density of specified numerical values of the reaction coordinates, $( \xi,r^\alpha)$, given by
\begin{equation}
P( \xi,r^\alpha)=\langle \delta(\xi_i(\bm{r}_{i}^{n_a})-\xi)\delta(r^\alpha_{i}(\bm{r}_{i}^{n_a}))-r^\alpha)\rangle_t,
\end{equation}
where the angle brackets denote an average over the local nonequilibrium distribution defined below (Eq.~(\ref{eq:local_eq-new})) and $P_u$ is a uniform probability density. The free energy $W(\xi,r^\alpha)$ has the form shown schematically in Fig.~\ref{fig:W}.

The species variables may be defined in terms of $\xi_i(\bm{r}_i^{n_r})$ as
\begin{equation}\label{eq:spexies-variables}
\theta^\gamma_i(\xi_i)=\Theta_i^R H_\gamma(\xi_i(\bm{r}_i^{n_r})),
\end{equation}
where $H_\gamma(\xi_i(\bm{r}_i^{n_r}))$ restricts molecular configurations to species $\gamma \in \{A,B\}$: $H_A(\xi_i(\bm{r}_i^{n_r}))=H(\xi^\ddagger-\xi_i(\bm{r}_i^{n_r}))$ and $H_B(\xi_i(\bm{r}_i^{n_r}))=H(\xi_i(\bm{r}_i^{n_r})-\xi^\ddagger)$ with $H$ a Heaviside function. The local number density of reactive molecules at a field point $\bm{r}$ with origin at the center of the colloid is given by
\begin{equation}
N_R(\bm{r})= \sum_{i=1}^N \Theta^R_i \delta(\bm{r}_{ic}-\bm{r}),
\end{equation}
and it can be partitioned into the sum of the local number densities of the $A$ and $B$ species at this field point, $N_R(\bm{r})=N_A(\bm{r})+N_B(\bm{r})$, where
\begin{equation}\label{eq:A-B-densities}
N_\gamma(\bm{r})= \sum_{i=1}^N \theta^\gamma_i(\xi_i) \delta(\bm{r}_{ic}-\bm{r}).
\end{equation}
These densities are important quantities for the specification of the nonequilibrium state of the system and enter the reaction-diffusion equation for the system.
The fluxes of these species densities in the presence of a fixed colloid are given by
\begin{equation}\label{eq:Ngamma-flux}
\dot{N}_{\gamma{}}(\bm{r}) = i\Liou_0 N_{\gamma}(\bm{r}) =  J_\gamma^R(\bm{r})-\bm{\nabla{}}_r\cdot{}\bm{j}_{\gamma}(\bm{r}),
\end{equation}
where the local reaction rate and the number density fluxes of species $\gamma$ are
\begin{eqnarray}\label{eq:rate-flux}
J_\gamma^R(\bm{r})&=&\sum_{i=1}^N \dot{\theta}_i^{\gamma}(\xi_i)\delta(\bm{r}_{ic}-\bm{r}), \\ \bm{j}_{\gamma}(\bm{r})&=&\sum_{i=1}^{N}\Theta_{i}^{\gamma{}}m^{-1}\bm{p}_{i}
 \delta (\bm{r}_{ic} -\bm{r}).
\end{eqnarray}

\section{Nonequilibrium state of the fluid} \label{sec:nonequilibrium}

In the microscopic theory of Brownian motion in an equilibrium system developed by Mazur and Oppenheim~\cite{MO70} the Langevin equation is obtained by applying a projection operator formalism in which the projection extracts the average of dynamical variables over the equilibrium bath density $\rho_e$ in the presence of a fixed colloid. This density is stationary under the Liouville operator $i\Liou_0$ for a system in which the colloid is held fixed.

However, active motion can take place only under nonequilibrium conditions and the constraints that drive the system out of equilibrium must be specified. If the system is maintained out of equilibrium by an externally-imposed reservoir, the entropy production of the fluid is nonzero at all times and the bath density $\rho_b$  does not equilibrate to $\rho_e$ but instead evolves according to the Liouville equation for the bath in the presence of a fixed colloid,
\begin{equation}\label{eq:liou-bath}
\partial_t \rho_b(t)=-i\Liou_0\rho_b(t).
\end{equation}

To implement the constraints imposed by the external reservoirs we use the statistical mechanical theory for transport processes in systems out of equilibrium.~\cite{R67,P68,OL79,SO96,SO97,SO98,CDDEDC18,CDDCE19}
The nonequilibrium state of the fluid is determined by a set of conjugate fields that couple to the following local fluid fields: the number density $N_\gamma (\bm{r})$ of the reactive solute species defined in Eq.~(\ref{eq:A-B-densities}), the total number density of fluid molecules,
\begin{equation}\label{eq:number-density}
N(\bm{r}) = \sum_{i=1}^{N}\delta(\bm{r}_{ic}-\bm{r}),
\end{equation}
that is equal to the sum of solvent and solute densities, $N(\bm{r})=N_S(\bm{r})+N_A(\bm{r})+N_B(\bm{r})$, where the solvent density is
\begin{equation}\label{eq:solvent-density}
N_S(\bm{r})= \sum_{i=1}^N \Theta^S_i \delta(\bm{r}_{ic}-\bm{r}),
\end{equation}
the total momentum density of the centers of mass of the solvent and solute molecules,
\begin{equation}\label{eq:momentum-density}
\bm{g}_N(\bm{r})=
\sum_{i=1}^{N}\bm{p}_{i}\delta(\bm{r}_{ic}-\bm{r})
\end{equation}
and the energy density of the fluid particles in the presence of the colloid
\begin{equation}\label{eq:energy-density}
E_N(\bm{r})=\sum_{i=1}^N \Big[\Theta_i^S \frac{p_{i}^{2}}{2m} +\Theta_i^R H_{mi} +U_{{\rm f}i}+U_{{\rm I}i}\Big] \delta(\bm{r}_{ic}-\bm{r}).
\end{equation}

 We further assume that the system is isothermal with temperature $T$ although the formulation can be generalized to accommodate temperature variations. Note that the constraints are applied to the species densities $N_\gamma (\bm{r})$ and  total number and momentum densities. They are not applied to the total reactive molecule density $N_R(\bm{r})$ since we are primarily interested in situations where the species densities are maintained out of equilibrium. We then consider the set of fluid fields,
\begin{equation}\label{eq:A-fields}
\bm{A}(\bm{r})=\{N_{\gamma}(\bm{r}),N(\bm{r}),\bm{g}_N(\bm{r}),E_N(\bm{r})\}.
\end{equation}
and corresponding conjugate fields,
\begin{eqnarray}
&&\bm{\phi}_{A}(\bm{r},t)=\\
&&\quad \{ \beta \tilde{\mu}_{\gamma}(\bm{r},t), \beta({\mu}_{S}(\bm{r},t) - \frac{1}{2}mv^{2}(\bm{r},t)),\beta\bm{v}(\bm{r},t), -\beta \},\nonumber
\end{eqnarray}
where $\beta=1/(k_{B}T)$ with $k_{B}$ Boltzmann's constant. The local relative chemical potential of species $\gamma$ is $\tilde{\mu}_{\gamma{}}(\bm{r},t)=\mu_{\gamma{}}(\bm{r},t)-\mu_{S}(\bm{r},t)$ while $\bm{v}(\bm{r},t)$ is the local fluid velocity field. The approach can be generalized to include mode coupling contributions by expanding the set of variables to include all nonlinear products of the slowly-varying fields.~\cite{KO88,SLO92}

The local nonequilibrium distribution function may be written as
\begin{equation}\label{eq:local_eq-new}
\rho_{L}(t)=\frac{\prod_\lambda (N_\lambda ! h^{3N_\lambda})^{-1}e^{\bm{A}(\bm{r}) \ast{} \bm{\phi}_{A}(\bm{r},t)}}{\Tr[
\prod_\lambda (N_\lambda ! h^{3N_\lambda})^{-1}e^{\bm{A}(\bm{r})\ast \bm{\phi}_{A}(\bm{r},t)}]},
\end{equation}
where $\ast{}$ denotes a scalar product and an integration over $\bm{r}$, i.e., $\bm{A}(\bm{r})\ast{}\bm{\phi}_{A}(\bm{r},t)=\int{}d\bm{r}\bm{A}(\bm{r})\cdot{}\bm{\phi}_{A}(\bm{r},t)$, and $\lambda \in \{S,A,B\}$. The trace operation includes an integration over phase space and a sum over particle numbers and types,
\begin{equation}\label{eq:trace}
\Tr[\cdots]=\prod_\lambda \sum_{N_{\lambda}=0}^{\infty{}}\int{}d\bm{x}^{N_S} d\bm{x}_m^{N_R}\; \cdots.
\end{equation}

The values of the conjugate fields $\bm{\phi}_{A}(\bm{r},t)$ are chosen such that the local nonequilibrium averages of the $\bm{A}(\bm{r})$ variables in the presence of the colloid are given by their exact nonequilibrium averages in the presence of a fixed colloid,
\begin{equation}\label{eq:constraint-cond}
\bm{a}(\bm{r},t) \equiv \Tr[\rho_{b}(t)\bm{A}(\bm{r})]=\Tr[\rho_{L}(t)\bm{A}(\bm{r})]\equiv \langle{}\bm{A}(\bm{r})\rangle{}_{t}.
\end{equation}
Both $\rho_b(t)$ and $\rho_L(t)$ depend parametrically on the fixed position $\bm{R}$ and orientation $\bm{\theta}$ of the colloidal particle, explicitly through the interaction potential in the Hamiltonian and through the thermodynamic conjugate fields $\phi_A(\bm{r},t)$.

The local equilibrium distribution function $\rho_L(t)$ can be generalized to incorporate additional higher-order conjugate fields that couple to nonlinear products of the hydrodynamic densities. The additional conjugate fields are important when considering the dynamics of multilinear densities in nonequilibrium systems where the densities can exhibit long range correlations.  However for linear densities of hydrodynamic fluid fields, the additional conjugate fields provide only small mode-coupling corrections that can be neglected to a good approximation.~\cite{SO94}

To study the self-diffusiophoretic motion of the colloid, the solute chemical potentials can be given specified values far from the particle to describe a nonequilibrium scenario in which fuel and product species are fed in or removed from the system using external reservoirs. In this circumstance the fluid velocity field vanishes far from the colloid and there are no net fluid flows, although fluid flows are produced in the vicinity of the colloid as part of the diffusiophoretic mechanism.

\section{Derivation of generalized Langevin equation}\label{sec:LangevinDerivation}

Preparatory to obtaining the equations of motion for the linear and angular momenta of the colloid, we first consider how the Langevin equation for a general function $\bm{D}(\bm{X})$ of the colloidal degrees of freedom may be obtained. The variable $\bm{D}(\bm{X},t)$ satisfies the equation of motion,
\begin{equation} \label{eq:D}
\frac{d}{dt}\bm{D}(t) = i\Liou \bm{D}(t)= e^{i\Liou t} i\Liou \bm{D}(0).
\end{equation}
The generalized Langevin equation is obtained from Eq.~(\ref{eq:D}) by projecting out the bath degrees of freedom so that their effects are incorporated in frictional and random forces. In order to project out the dependence on the bath variables we make use of the time-dependent projector $\mathcal{P}(t)$ defined by its action on an arbitrary function $f$,~\cite{OL79,SO96}
\begin{equation}
\mathcal{P}(t) f = \Tr [ \rho_b(t) f],
\end{equation}
and its complement, $\mathcal{Q}(t) = 1 - \mathcal{P}(t)$. The adjoint of the projector $\mathcal{P}(t)$ is $\mathcal{P}^\dagger(t)$ defined by
$\mathcal{P}^\dagger(t) f = \rho_b(t)  \Tr [f]$. Following usual methods, the generalized Langevin equation is obtained by rewriting the propagator $U(0,t)=\exp{(i\Liou{} t)}$ in an equivalent form involving the time-ordered projected propagator $U_Q(0,t)=\mathcal{T}_- \exp{(\int_0^t dt_1 \; i\Liou \mathcal{Q}(t_1))}$ where $\mathcal{T}_-$ is a time ordering operator that orders operators in increasing order of their time argument. As shown in Appendix~\ref{sec:evolution-operator} the evolution operators $U(0,t)$ and $U_Q(0,t)$ are related by
\begin{eqnarray} \label{eq:U-UQ-relation}
U(0,t)&=&  U(0,t)\mathcal{P}(t) + \mathcal{Q}(0)U_Q(0,t) \\
&-&\int_0^t dt_1 \; U(0,t_1) (\partial_{t_1}\mathcal{P}(t_1)) U_Q(t_1,t)\nonumber \\
&+&\int_0^t dt_1 \; U(0,t_1) \mathcal{P}(t_1) i\Liou{} \mathcal{Q}(t_1) U_Q(t_1,t).\nonumber
\end{eqnarray}

Inserting this expression for $U(0,t)=\exp{(i\Liou{} t)}$ into the equation of motion~(\ref{eq:D}), we obtain
\begin{eqnarray}\label{eq:PLang1}
\frac{d}{dt} \bm{D}(t)&=&  e^{i\Liou{} t}\mathcal{P}(t)\dot{\bm{D}} + \bm{F}^{\rm D}_{\rm fl}(t) \nonumber \\
&-& \int_0^t dt_1 \; e^{i\Liou{} t_1} (\partial_{t_1}\mathcal{P}(t_1)) \bm{K}_D(t_1,t) \nonumber \\
&+& \int_0^t dt_1 \; e^{i\Liou{} t_1} \mathcal{P}(t_1) i\Liou{} \bm{K}_D(t_1,t),
\end{eqnarray}
where we have defined
\begin{eqnarray}\label{eq:K}
\bm{K}_D(t_1,t_2)&=& \mathcal{Q}(t_1)U_Q(t_1,t_2)\dot{\bm{D}} \nonumber\\
&=& \mathcal{Q}(t_1)U_Q(t_1,t_2)\mathcal{Q}(t_2)\dot{\bm{D}},
\end{eqnarray}
and made use of the relation $\mathcal{P}(t_1) + \mathcal{Q}(t_1) =1$ in writing the third term on the right of Eq.~(\ref{eq:PLang1}). The fluctuating force is given by $\bm{F}^{\rm D}_{\rm fl}(t)=\bm{K}_D(0,t)$.

The integral terms in Eq.~(\ref{eq:PLang1}) can be evaluated as shown in Appendix~\ref{sec:integral-terms} and using these results the generalized Langevin equation for $\bm{D}(t)$ reads
\begin{eqnarray}\label{eq:PLang1sup}
&&\frac{d}{dt} \bm{D}(t)=  \Tr[\rho_b(t)\dot{\bm{D}}](\bm{X}(t)) + \bm{F}^{\rm D}_{\rm fl}(t) \\
&+&  \int_0^t dt_1 \;  \Big( -\frac{\bm{P}(t_1)}{M} \cdot \bm{M}_1(t_1,t) + \bm{\nabla}_{\bm{P}(t_1)} \cdot \bm{M}_2(t_1,t)\nonumber \\
&-&  \bm{\Pi}(t_1)^T \cdot {\bm{M}(t_1)}^{-1}\cdot \bm{M}_3(t_1,t) - \bm{\nabla}_{\bm{\Pi}(t_1)} \cdot \bm{M}_4(t_1,t) \Big),\nonumber
\end{eqnarray}
where we defined
\begin{eqnarray}\label{eq:M-quantities}
\bm{M}_1(t_1,t; \bm{X})&=&\Tr [(\bm{\nabla}_R \rho_b(t_1))  \bm{K}_D(t_1,t)],\\
\bm{M}_2(t_1,t; \bm{X})&=&\Tr [\rho_b(t_1)\bm{F}_c \bm{K}_D(t_1,t)],\\
\bm{M}_3(t_1,t; \bm{X})&=&\Tr [(\bm{\nabla}_\theta \rho_b(t_1))  \bm{K}_D(t_1,t)],\\
\bm{M}_4(t_1,t; \bm{X})&=& \Tr [\rho_b(t_1)) \bm{\nabla}_\theta U_I \bm{K}_D(t_1,t)],
\end{eqnarray}
but have not indicated the dependence of these quantities on $\bm{X}(t_1)$ in Eq.~(\ref{eq:PLang1sup}). The matrix $\bm{M}(t_1)$ in Eq.~(\ref{eq:PLang1sup}) corresponds to the mass-weighted kinetic matrix $\bm{M}$ defined by Eq.~(\ref{eq:rotationalKE}) evaluated at the fixed colloid position and orientation at time $t_1$.

\subsection{Approximate form of Langevin equation: Brownian motion scaling}
Following the theory of Brownian motion~\cite{MO70}, when $M\gg m$ it is useful to introduce scaled momenta, $\bm{P}^*=\mu \bm{P}$ and $\bm{\Pi}^* = \mu \bm{\Pi}$, where $\mu=(m/M)^{1/2}$ is a small parameter that gauges the magnitude of the colloidal momenta. The corresponding scaled colloidal Liouvillian is $i\Liou{}_c=\mu i\Liou{}_c^*$. The above results, along with the action of $\exp({i \Liou t})$,  allow us to write the generalized Langevin equation for the colloid in scaled colloidal coordinates as
\begin{eqnarray}\label{eq:PLang2}
&&\frac{d\bm{D}^*(t)}{dt} =  \mu \Tr [\rho_b(t)\dot{\bm{D}}](\bm{X}(t)) +\mu \bm{F}^{D}_{\rm fl}(t) + \\
&\mu^2& \int_0^t dt_1 \;  \Big( -\frac{\bm{P}^*(t_1)}{m} \cdot \bm{M}_1(t_1,t) + \bm{\nabla}_{P^*(t_1)} \cdot \bm{M}_2(t_1,t)\nonumber \\
&-& \bm{\Pi}^*(t_1) \cdot {\bm{M}(t_1)}^{-1} \cdot \bm{M}_3(t_1,t)-\bm{\nabla}_{\Pi^*} (t_1)\cdot \bm{M}_4(t_1,t)\Big).\nonumber
\end{eqnarray}
Again, we have not indicated the dependence of the matrices of transport coefficients $\bm{M}_{i}$ on $\bm{X}$ for simplicity.

The relation between the exact and local nonequilibrium distributions, $\rho_b(t)$ and $\rho_L(t)$, respectively, is given in Eq.~(\ref{eq:rhob-rhoL2}). The fields, $\mu_{\lambda}(\bm{r},t)$ and $\bm{v}(\bm{r},t)$ in this equation are assumed to be slowly varying in space so we may associate a small parameter $\epsilon_h$ that gauges the size of the gradients of these fields. In addition we assume that the reactions are rare events and associate another small parameter $\epsilon_r$ that gauges the magnitude of the reactive flux. Equation~(\ref{eq:rhob-rhoL2}) also contains a term $\mathcal{Q}_A(t_1)\bm{F}_{\rm f}(\bm{r}) \ast  \bm{v}(\bm{r},t_1)$. The $\mathcal{Q}_A$ projector removes the contributions to $\bm{F}_{\rm f}(\bm{r})$ that are proportional to the species and total number densities, leaving only contributions to the force that arise from internal molecular degrees of freedom. Neglecting such contributions we have $\rho_b(t) =\rho_L(t)+ {\mathcal O}(\epsilon_{h,r})$ and we can replace $\rho_b(t)$ by $\rho_L(t)$ in evaluating the $\bm{M}_{i}$ functions. However, since $\Tr [\rho_b(t)\dot{\bm{D}}]$ scales as $\mu$, this replacement cannot be made in this term.

Since $\bm{\nabla}_R \rho_L(t) = \beta \left( \bm{F}_c - \langle \bm{F}_c \rangle_t \right) \rho_L(t)$ and $\bm{\nabla}_\theta \rho_L(t) = -\beta \left( \bm{\nabla}_\theta U_I - \langle \bm{\nabla}_\theta U_I \rangle_t \right) \rho_L(t)$, in this approximation we have
\begin{eqnarray}\label{eq:K2}
\bm{M}_1(t_1,t) &=& \Tr [(\bm{\nabla}_R \rho_L(t_1))  \bm{K}_D(t_1,t)] \nonumber\\
&=&\beta \Tr [\rho_L(t_1)) \bm{F}_c \bm{K}_D(t_1,t)]\nonumber\\
&=&\beta \bm{M}_2(t_1,t),\\
\bm{M}_3(t_1,t) &=& \Tr [(\bm{\nabla}_\theta \rho_L(t_1))  \bm{K}_D(t_1,t)] \nonumber\\
&=& -\beta \Tr [\rho_L(t_1)) \bm{\nabla}_\theta U_I \bm{K}_D(t_1,t)]\nonumber\\
&=&-\beta \bm{M}_4(t_1,t).
\end{eqnarray}
Noting that $\bm{\nabla}_\theta U_I = - \bm{N} \cdot \bm{T}$ and $\bm{\Pi}^T \cdot \bm{M}^{-1} \bm{\nabla}_\theta U_I = -\bm{L}\cdot \bm{I_m}^{-1} \cdot \bm{T}$, the last two terms of Eq.~(\ref{eq:PLang2}) can be written in terms of the scaled angular momentum $\bm{L}^*$ as
\begin{eqnarray}
&&-\bm{\Pi}^*(t_1) \cdot \bm{M}(t_1)^{-1} \cdot \bm{M}_3(t_1,t) - \bm{\nabla}_{\Pi^* (t_1)} \cdot \bm{M}_4(t_1,t) =\nonumber \\
&&\Big( - \beta \bm{L}^* (t_1) \cdot {\bm{I_m} (t_1)}^{-1}  + \bm{\nabla}_{\bm L^*(t_1)} \Big)
 \cdot \bm{M}_{T\dot{D}}(t_1,t),
\end{eqnarray}
where $\bm{M}_{T\dot{D}}(t_1,t) = \Tr [\rho_L(t_1) \bm{T} \bm{K}_{D}(t_1,t)]$. We also let $\bm{M}_{F\dot{D}}(t_1,t) = \bm{M}_{2}(t_1,t)$.

Using the equations given above we can write $\bm{M}_{FD}$ and $\bm{M}_{TD}$ more explicitly in the form of friction kernels:
\begin{eqnarray}\label{eq:M_FTD}
\bm{M}_{F\dot{D}}(t_1,t) &=& \langle (\bm{F}_c - \langle \bm{F}_c \rangle_{t})\mathcal{Q}(t_1)U_{Q}(t_1,t)(\dot{\bm{D}} - \langle \dot{\bm{D}} \rangle_{t}) \rangle_{t_1} \nonumber \\
\bm{M}_{T\dot{D} }(t_1,t) &=& \langle (\bm{T} - \langle \bm{T}\rangle_t) \mathcal{Q}(t_1)U_{Q}(t_1,t)(\dot{\bm{D}} - \langle \dot{\bm{D}} \rangle_{t}) \rangle_{t_1},\nonumber\\
\end{eqnarray}
where $U_{Q}(t_1,t)$ is now taken to be the projected evolution operator with $\rho_b$ replaced by $\rho_L$ and $i\Liou$ by $i\Liou_0$ neglecting higher order $\mu$ contributions.

\section{Langevin equations for linear and angular momenta}\label{sec:linear-angular}

Taking the dynamical variables $\bm{D} = \bm{P}$ and $\bm{D}=\bm{L}$, noting the fact that $U_{Q}(t_1,t) \approx e^{i\Liou_0 (t-t_1)} \left( 1 + {\mathcal O}(\mu) + {\mathcal O}(\epsilon_{h,r})\right)$  and retaining only the lowest order terms of the small parameters $\mu$, $\epsilon_h$ and $\epsilon_r$ in Eq.~(\ref{eq:M_FTD}), we obtain the coupled Langevin equations for translational and rotational motion of the colloid in the unscaled coordinates,
\begin{eqnarray}
&&\frac{d}{dt}\bm{P}(t) =  \Tr [\rho_b(t)\bm{F}_c ] \big( \bm{R}(t), \bm{\theta}(t) \big) + \bm{F}_{\rm fl}(t) \\,
&& \quad -  \int_0^t dt_1 \; \beta \frac{\bm{P}(t_1)}{M} \cdot \bm{M}_{FF}(t_1,t) \nonumber \\
&& \quad -  \int_0^t dt_1 \; \beta \bm{L}(t_1) \cdot {\bm{I_m}(t_1)}^{-1} \cdot \bm{M}_{TF}(t_1,t),\nonumber
\end{eqnarray}
and
\begin{eqnarray}
&&\frac{d}{dt}\bm{L}(t) =  \Tr [\rho_b(t)\bm{T}]\big(\bm{R}(t), \bm{\theta}(t) \big) + \bm{T}_{\rm fl}(t) \\
&& \quad-  \int_0^t dt_1 \; \beta \frac{\bm{P}(t_1)}{M}  \cdot \bm{M}_{FT}(t_1,t) \nonumber \\
&& \quad-  \int_0^t dt_1 \; \beta \bm{L}(t_1) \cdot {\bm{I_m} (t_1)}^{-1} \cdot \bm{M}_{TT}(t_1,t), \nonumber
\end{eqnarray}
where we used the notation $\bm{F}^{P}_{\rm fl}=\bm{F}_{\rm fl}$ and $\bm{F}^{L}_{\rm fl}=\bm{T}_{\rm fl}$ for the random force and torque.
The generalized Langevin equation for the linear momentum takes the form of an ordinary Langevin equation by changing variables $t'=t-t_1$, taking $\bm{P}(t-t') \approx \bm{P}(t)$  on the fast time scale of the force correlation decay, and defining the friction tensor by
\begin{eqnarray}
\mbox{\boldmath{$\zeta$}}_t &=& \beta \int_0^\infty dt' \; \bm{M}_{FF}(0,t') \\
&=&\beta \int_0^\infty dt' \;\langle (\bm{F}_c - \langle \bm{F}_c \rangle_{t}) e^{i\Liou_0 t'}(\bm{F}_c - \langle \bm{F}_c \rangle_{t}) \rangle_{t}.\nonumber
\end{eqnarray}
In a similar approximation, the rotational friction tensor $\mbox{\boldmath{$\zeta$}}_r$ can be defined as
\begin{eqnarray}
\mbox{\boldmath{$\zeta$}}_r &=& \beta \int_0^\infty dt' \; {\bm{M}}_{TT}(0,t')  \\
&=&\beta \int_0^\infty dt' \;\langle (\bm{T} - \langle \bm{T} \rangle_{t}) e^{i\Liou_0 t'}(\bm{T} - \langle \bm{T} \rangle_{t}) \rangle_{t},\nonumber
\end{eqnarray}
with analogous expressions for the cross-coupling friction tensors $\mbox{\boldmath{$\zeta$}}_{tr}$ and $\mbox{\boldmath{$\zeta$}}_{rt}$  that couple translational and rotational motion.

Setting $\bm{P}=M \bm{V}$, when the translational and rotational motion decouple the Langevin equations take the final form~\cite{mass-scaling}
\begin{eqnarray}\label{eq:PLangF}
M\frac{d}{dt}\bm{V}(t) &=&  \Tr [\rho_b(t)\bm{F}_c] \big(\bm{R}(t), \bm{\theta} (t) \big)  \nonumber \\
&&\qquad - \zeta_t  \bm{V}(t) + \bm{F}_{\rm fl}(t), \\
\frac{d\bm{L}(t)}{dt} &=& \Tr [\rho_b(t)\bm{T}] \big(\bm{R}(t), \bm{\theta}(t) \big) \nonumber \\
&& \qquad - \zeta_r  \bm{L}(t) \cdot {\bm{I_m}(t)}^{-1}+ \bm{T}_{\rm fl}(t),
\end{eqnarray}
where we used $\mbox{\boldmath{$\zeta$}}_t=\zeta_t \bm{1}$ and $\mbox{\boldmath{$\zeta$}}_r=\zeta_r \bm{1}$.
For a spherical rotor, ${\bm{I_m}}(t) = I_1 {\bm{U}}$ is diagonal and independent of time. In this case the angular momentum is ${\bm{L}}(t) = I_1 \bm{\omega} (t)$ and an equation of Langevin form can be written for the angular velocity $\bm{\omega}(t)$.

\section{Diffusiophoretic force and torque}\label{sec:averageF}

The the mean force and torque in the Langevin equations, $\Tr [\rho_b(t)\bm{F}_c](\bm{R}(t), \bm{\theta}(t))$ and $\Tr [\rho_b(t)\bm{T}](\bm{R}(t),\bm{\theta}(t))$, respectively,  are responsible for the active translational and rotational motion of the colloid. In the absence of constraints that drive the system out of equilibrium both of these quantities vanish and Langevin equations reduce the standard forms that describe the Brownian dynamics of inactive colloids.

The constraints described by the $\bm{\phi}_{A}(\bm{r},t)$ fields can be applied in various ways to specify the nonequilibrium state. For a self-diffusiophoretic colloid a simple constraint is the specification of the values of the $A$ and $B$ species chemical potentials far from the colloid. To study more general aspects of diffusiophoretic colloidal motion, the gradients of these chemical potentials could also be specified. Under such constraints the fluid velocity fields vanish far from the colloid, although, as noted earlier,  the active motion of the colloid will generate local variations of the concentration and velocity fields in the vicinity of the colloid as part of the diffusiophoretic mechanism. In this section we consider the forms that the diffusiophoretic force and torque take under such constraints.

\subsection{Force}
Using momentum conservation the force on the colloid can be written in terms of the local force on the fluid, $\bm{F}_c = -\int d\bm{r} \;\bm{F}_{\rm f}(\bm{r})$, given in Eq.~(\ref{eq:force-on-fluid}), as
\begin{eqnarray}\label{eq:colloid-force-tot}
&&\bm{F}_c =  \sum_{\alpha =1}^{n_s} \Big[\int d\bm{r} \;\sum_{i=1}^N\sum_{b=N}^C \Theta_\alpha^b\Big(\Theta_i^S \bm{\nabla_{r_i}} V_{Sb}(r_i^\alpha)  \\
&+&\Theta_i^R \sum_{k=1}^{n_a}\bm{\nabla_{r_{(k)i}}}V_{kb}(r_{(k)i}^{\alpha})
\Big)\delta(\bm{r}_{ic}-\bm{r})\Big] \equiv \sum_{\alpha =1}^{n_s} \bm{F}_c^\alpha,\nonumber
\end{eqnarray}
and $\bm{F}_c^\alpha$ can be written terms of the local solvent and $\bm{r}^{n_a}$-dependent reactive molecule densities as,
\begin{eqnarray}\label{eq:colloid-force-alpha-den}
&&\bm{F}_c^\alpha = \int d\bm{r} \;  \sum_{b=N}^C \Theta_\alpha^b \big(\bm{\nabla_r} V_{Sb}(r^\alpha)\big) N_S(\bm{r}) + \\
&&\int d\bm{r} \; \int d\bm{r}^{n_a} \;   \sum_{b=N}^C \Theta_\alpha^b \Big[ \sum_{k=1}^{n_a}\bm{\nabla_{r_{(k)}}}V_{kb}(r_{(k)}^{\alpha}) \Big]N_R(\bm{r},\bm{r}^{n_a}).\nonumber
\end{eqnarray}
The local $\bm{r}^{n_a}$-dependent reactive molecule density is defined by
\begin{equation}
N_R(\bm{r},\bm{r}^{n_a})= \sum_{i=1}^N \Theta^R_i \delta(\bm{r}_{ic}-\bm{r}) \delta(\bm{r}^{n_a}_i-\bm{r}^{n_a}).
\end{equation}

The expression for the force on the colloid in Eq.~(\ref{eq:colloid-force-alpha-den}) involves $N_R(\bm{r},\bm{r}^{n_a})$ and not the $A$ and $B$ species densities that enter the constraint conditions in Eq.~(\ref{eq:constraint-cond}). We can rewrite it in terms of $N_\gamma(\bm{r})$ using projectors that project $N_R(\bm{r},\bm{r}^{n_a})$ onto the species densities. We let $p(\bm{r}^{n_a}|r)$ be the conditional probability density of the molecular coordinates $\bm{r}^{n_a}$ given a distance $r$ of the center of mass of the molecule from the colloid center, and define a projector ${\sf P}_s$ and its complement ${\sf Q}_s=1-{\sf P}_s$ by
\begin{eqnarray}
{\sf P}_s f(\bm{r}^{n_a}) &=& \sum_\gamma p_\gamma(\bm{r}^{n_a}|r) \int d\bm{r}^{n_a} \; H_\gamma(\xi(\bm{r}^{n_a})) f(\bm{r}^{n_a}) \nonumber \\
&\equiv& \sum_\gamma{\sf P}^\gamma_s f(\bm{r}^{n_a}) ,
\end{eqnarray}
where
\begin{equation}
p_\gamma(\bm{r}^{n_a}|r)=\frac{H_\gamma(\xi(\bm{r}^{n_a})) p(\bm{r}^{n_a}|r)}{\int d\bm{r}^{n_a}\; H_\gamma(\xi(\bm{r}^{n_a})) p(\bm{r}^{n_a}|r)}
\end{equation}
is the conditional probability density with the  internal molecular coordinates restricted to those for species $\gamma$. The action of this projector on $N_R(\bm{r},\bm{r}^{n_a})$ is
\begin{equation}
{\sf P}_s N_R(\bm{r},\bm{r}^{n_a})= \sum_\gamma p_\gamma(\bm{r}^{n_a}|r)N_\gamma(\bm{r}).
\end{equation}
Inserting $N_R(\bm{r},\bm{r}^{n_a})= {\sf P}_s N_R(\bm{r},\bm{r}^{n_a})+{\sf Q}_s N_R(\bm{r},\bm{r}^{n_a})$  in Eq.~(\ref{eq:colloid-force-alpha-den}) we get
\begin{equation}
\bm{F}_c^\alpha = \sum_\lambda \int d\bm{r} \; N_\lambda(\bm{r})
\sum_{b=N}^C \bm{F}_{\lambda b}^\alpha (\bm{r}) + \Delta \bm{F}_c^\alpha,
\label{siteForce}
\end{equation}
where $\bm{F}_{Sb}^\alpha=\Theta_\alpha^b\bm{\nabla_r}V_{Sb}(r^\alpha)$ and
\begin{eqnarray}
&&\bm{F}_{\gamma b}^\alpha (\bm{r}) = \Theta_{\alpha}^b \int d\bm{r}^{n_a} \;
\bigg[ \sum_{k=1}^{n_a}\bm{\nabla_{r_{(k)}}}V_{kb}(r_{(k)}^{\alpha}) \bigg] p_\gamma(\bm{r}^{n_a} | \bm{r}), \\
&&\Delta \bm{F}_c^\alpha = \Theta_\alpha^b
\int d\bm{r} \, d\bm{r}^{n_a} \; \bigg[ \sum_{k=1}^{n_a}\bm{\nabla_{r_{(k)}}}V_{kb}(r_{(k)}^{\alpha})\bigg] {\sf Q}_s N_R(\bm{r}, \bm{r}^{n_a}).\nonumber\\
\end{eqnarray}

Using $\langle N(\bm{r}) \rangle_t= \sum_\lambda \langle N_\lambda(\bm{r}) \rangle_t$, and the notation introduced in Eq.~(\ref{eq:constraint-cond}) where $\langle N(\bm{r}) \rangle_t=n(\bm{r},t)$ and $\langle N_\gamma(\bm{r}) \rangle_t=n_\gamma(\bm{r},t)$, the diffusiophoretic force may now be written as
\begin{eqnarray}\label{eq:diffusiophoretic-force-colloid}
&&\Tr [\rho_b(t)\bm{F}_c] \big( \bm{R}(t), \bm{\theta}(t) \big) =\\
&&\int d\bm{r} \;\Big[\sum_{\gamma} \Big(\sum_{b=N}^C  \big(\bm{F}_{\gamma b}(\bm{r})-\bm{F}_{S b}(\bm{r})\big)\Big) n_\gamma(\bm{r},t) \nonumber \\
&& \qquad + \Big(\sum_{\lambda} \sum_{b=N}^C  \bm{F}_{\lambda b}(\bm{r})\Big) n(\bm{r},t) \Big] + \Tr [\rho_b(t)\Delta \bm{F}_c ].\nonumber
\end{eqnarray}

\subsection{Torque}
A similar calculation can be carried out for the torque starting from the expression
\begin{eqnarray}
\bm{T} &=& - \bm{N}^{-1} \cdot \nabla_{\bm{\theta}} \sum_{\alpha =1}^{n_s} U_I^\alpha \nonumber \\
&=& - \sum_{\alpha =1}^{n_s} \bm{N}^{-1} \cdot \nabla_{\bm{\theta}} U_I^\alpha = \sum_{\alpha =1}^{n_s} \bm{T}^\alpha,
\label{Torque}
\end{eqnarray}
where $\bm{T}^\alpha$ is the contribution to the total torque from interaction site $\alpha$ on the colloid.  Noting that the $\bm{\theta}$-dependence of the interaction potential arises from the relative position $\bm{S}^\alpha(\bm{R}) =  \bm{S}^\alpha -\bm{R}= \bm{A}^T(\bm{\theta}) \cdot \tilde{\bm{S}}^\alpha$ of the interaction site from the center of the colloid, we have
\begin{eqnarray}
\bm{T}^\alpha &=& - \bm{N}^{-1} \cdot \sum_{b=N}^C \Theta_\alpha^b \sum_{i=1}^N  \bigg[ \Theta_i^S \nabla_{\bm{\theta}} \bm{r}_{i}^\alpha
\cdot \bm{\nabla_{r_{i}^\alpha}} V_{sb}(\bm{r}_{i}^\alpha) \nonumber \\
&& +
\Theta_i^R \sum_{k=1}^{n_a} \nabla_{\bm{\theta}} \bm{r}_{(k)i}^{\alpha } \cdot \bm{\nabla_{r_{(k)i}^{\alpha }}} V_{kb} (\bm{r}_{(k)i}^{\alpha }) \bigg]
\nonumber \\
&=& \bm{N}^{-1} \cdot \sum_{b=N}^C \Theta_\alpha^b \sum_{i=1}^N  \bigg[ \Theta_i^S \nabla_{\bm{\theta}} {\bm{S}}^\alpha(\bm{R})
\cdot \bm{\nabla_{r_{i}^\alpha}} V_{sb}(\bm{r}_{i}^\alpha) \nonumber \\
&&+
\Theta_i^R \sum_{k=1}^{n_a} \nabla_{\bm{\theta}}  {\bm{S}}^\alpha(\bm{R}) \cdot \bm{\nabla_{r_{(k)i}^{\alpha}}} V_{kb} (\bm{r}_{(k)i}^{\alpha } ) \bigg] .
\label{siteTorque}
\end{eqnarray}
From the definition of the $\bm{N}$ matrix in Eq.~(\ref{Nmatrixdef}) with use of the identity $\epsilon^{ijb}\epsilon^{cdb} = \delta_{ic}\delta_{jd} - \delta_{id}\delta_{jc}$, we find that
\begin{eqnarray}
\epsilon^{ijb}{N}_{ab} &=& \frac{1}{2} \left(
{A}_{ei} \nabla_{\theta_a} {A}_{ej} - {A}_{ej} \nabla_{\theta_a} {A}_{ei} \right) \nonumber \\
&=& {A}_{ei} \, \nabla_{\theta_a} {A}_{ej},
\end{eqnarray}
since $\nabla_{\theta_a} \bm{A}^{T} \cdot \bm{A} = 0$.

Considering
\begin{align*}
(\bm{N}^{-1})_{ab} \nabla_{\theta_b} &S_c^\alpha(\bm{R})=
 (\bm{N}^{-1})_{ab} \nabla_{\theta_b} {A}_{dc} \tilde{S}_d^\alpha \\
  &=
-(\bm{N}^{-1})_{ab} {A}_{dc} \, \left( \nabla_{\theta_b} {A}_{de}
\right) \, S_{e}^\alpha(\bm{R}) ,
\end{align*}
and taking the relation above into account we get
\begin{eqnarray*}
(\bm{N}^{-1})_{ab} \nabla_{\theta_b} S_c^\alpha(\bm{R}) &=&
 - (\bm{N}^{-1})_{ab} \, \epsilon^{fce} {N}_{bf} \, S^\alpha_e (\bm{R}) \\
&=&  -\epsilon^{ace} S_e^\alpha (\bm{R}).
\end{eqnarray*}
Using this relation in Eq.~(\ref{siteTorque}), we find the simple result
\begin{eqnarray}
\bm{T}^\alpha &=&  \bm{S}^\alpha (\bm{R}) \wedge \bigg[  \sum_{i=1}^N \sum_{b=N}^C \Theta_\alpha^b \theta_i^S \nabla_{\bm{r}_{ic}^\alpha}  V_{sb}(\bm{r}_{ic}^\alpha) \nonumber \\
&&+
\theta_i^R \sum_{k=1}^{n_a} \nabla_{\bm{r}_{ic}^{k(\alpha )}} V_{kb} (\bm{r}_{ic}^{k(\alpha )}) \bigg] \nonumber \\
&=& \bm{S}^\alpha (\bm{R}) \wedge \bm{F}_c^\alpha.
\end{eqnarray}
The average of the diffusiophoretic torque then adopts a form that is analogous to that for the diffusiophoretic force,
\begin{eqnarray}\label{eq:diffusiophoretic-torque-colloid}
&&\Tr [\rho_b(t)\bm{T}] \big( \bm{R}(t), \bm{\theta}(t) \big) = \\
&&  \sum_{\alpha=1}^{n_s} \bm{S}^\alpha (t) \wedge \int d\bm{r} \Bigg[ \;\sum_{\gamma}  n_\gamma(\bm{r},t)\sum_{b=N}^C
\big(
\bm{F}^\alpha_{\gamma b}(\bm{r})-\bm{F}_{S b}^\alpha(\bm{r})
\big)
 \nonumber \\
&&\qquad+ n(\bm{r},t) \sum_{\lambda} \sum_{b=N}^C  \bm{F}^\alpha_{\lambda b} (\bm{r})
\Bigg] \nonumber \\
&&\qquad+ \sum_{\alpha =1}^{n_s} \bm{S}^\alpha (t) \wedge \Tr \big[\rho_b(t) \Delta \bm{F}^\alpha_c  \big]  \nonumber \\
&& = \sum_{\alpha = 1}^{n_s} \bm{S}^{\alpha} (t) \wedge \Tr [\rho_b(t)\bm{F}_c^\alpha ] \big( \bm{R}(t), \bm{\theta}(t) \big)  \nonumber ,
\end{eqnarray}
where $\bm{S}^\alpha (t) = \bm{S}^\alpha \big( \bm{R}(t), \bm{\theta}(t) \big)$ are the positions of the sites $\alpha$ relative to the center of the colloid at time $t$.

\subsection{Contributions to force and torque}
The diffusiophoretic force and torque in Eqs.~(\ref{eq:diffusiophoretic-force-colloid}) and (\ref{eq:diffusiophoretic-torque-colloid}) have several contributions. The first two contributions involve the local equilibrium averages of the species and total density fields whose values are fixed by the constraints to give the exact nonequilibrium values of these quantities. These average fields may be determined from the solutions of the generalized hydrodynamic equations they satisfy. The last terms still retain the averages over the exact nonequilibrium density.

The terms involving $\Tr [\rho_b(t)\Delta \bm{F}_c ]$ are expected to be small. While introduction of the species densities $N_\gamma(\bm{r})$ accounts for nonequilibrium effects through reaction, the projected microscopic reactive molecule density ${\sf Q}_s N_R(\bm{r},\bm{r}^{n_a})$ that enters $\Tr [\rho_b(t)\Delta \bm{F}_c ]$ accounts for a nonequilibrium in the internal molecular degrees of freedom induced by the reaction. While such nonequilibrium effects can be taken into account they are not a dominant effect and are expected to be small in most situations.

With these approximations $\Tr [\rho_b(t)\bm{F}_c] \approx \langle \bm{F}_c \rangle_t$ and $\Tr [\rho_b(t)\bm{T}] \approx \langle \bm{T} \rangle_t$ and the diffusiophoretic force and torque are given by ,
\begin{eqnarray}\label{eq:simple-force-torque}
&&\langle \bm{F}_c \rangle_t =\int d\bm{r} \;\Big[\Big(\sum_{\lambda} \sum_{b=N}^C  \bm{F}_{\lambda b}(\bm{r})\Big) n(\bm{r},t) \nonumber \\
&& \qquad + \sum_{\gamma} \Big(\sum_{b=N}^C  \big(\bm{F}_{\gamma b}(\bm{r})-\bm{F}_{S b}(\bm{r})\big)\Big) n_\gamma(\bm{r},t) \Big]   \\
&&\langle \bm{T} \rangle_t = \sum_{\alpha=1}^{n_s} \bm{S}^\alpha (t) \wedge \int d\bm{r} \Bigg[ n(\bm{r},t) \sum_{\lambda} \sum_{b=N}^C  \bm{F}^\alpha_{\lambda b} (\bm{r})
 \nonumber \\
&&\qquad+ \;\sum_{\gamma}  n_\gamma(\bm{r},t)\sum_{b=N}^C
\big(
\bm{F}^\alpha_{\gamma b}(\bm{r})-\bm{F}_{S b}^\alpha(\bm{r})
\big)
\Bigg] \nonumber \\
&& = \sum_{\alpha = 1}^{n_s} \bm{S}^{\alpha} (t) \wedge \langle \bm{F}_c^\alpha \rangle_t \big( \bm{R}(t), \bm{\theta}(t) \big),
\end{eqnarray}
and their evaluation requires a knowledge of the local nonequilibrium averages of the chemical species and total density fields, which we consider below.

\subsection{Reaction-diffusion equations for species densities}\label{sec:RDeq}

The generalized hydrodynamic equations for the nonequilibrium averages $\bm{a}(\bm{r},t)$ of slowly-varying densities $\bm{A}(\bm{r})$ of microscopic variables can be derived by noting that
\begin{eqnarray}
\partial_t \bm{a} (\bm{r},t)&=& \Tr [\partial_t \rho_b(t) \bm{A} (\bm{r}) ] = -\Tr \left[ \big( i\Liou_0 \rho_b(t) \big) \bm{A}(\bm{r}) \right] \nonumber \\
&=& \Tr [ \rho_b(t) i\Liou_0 \bm{A}(\bm(r))] = \Tr \left[ \rho_b(t) \dot{\bm{A}}(\bm{r}) \right].
\label{eq:gen-hydro-base}
\end{eqnarray}
Using the relation between $\rho_b(t)$ and the local equilibrium density $\rho_L(t)$ established in Appendix~\ref{sec:bathDensity}, the hydrodynamic equations assume the form
\begin{eqnarray}\label{eq:gen-Ahydro}
&&\partial_t \bm{a} (\bm{r},t)= \langle \dot{\bm{A}}(\bm{r}) \rangle_t
+f_{A,t}(\bm{r},t)\\
&-& \int_0^t dt_1 \; \langle \bm{\mathcal{F}}_{A,t}(\bm{r},t_1,t) \bm{\mathcal{F}}_{A,t_1}(\bm{r}')
\rangle_{t_1} \ast \bm{\phi}_A(\bm{r}',t_1),\nonumber
\end{eqnarray}
with
\begin{eqnarray}
\bm{\mathcal{F}}_{A,t}(\bm{r},t_1,t)&=&U_{Q_A}(t_1,t) \mathcal{Q}_{A}(t)i \mathcal{L}_0 \bm{A}(\bm{r})\nonumber \\
&=&U_{Q_A}(t_1,t) \bm{\mathcal{F}}_{A,t}(\bm{r}),
\end{eqnarray}
where
\begin{equation*}
U_{Q_A}(t_1,t) = {\mathcal T}_{-} \exp \left\{ \int_{t_1}^t dt_2 \,  {\mathcal Q}_A(t_2) i\Liou_0  \right\},
\end{equation*}
which follows by taking the Hermitian conjugate of Eq.~(\ref{eq:formalUdagger}),
and the random force is given by $f_{A,t}(\bm{r},t)=\Tr[\rho_b(0) \bm{\mathcal{F}}_{A,t}(\bm{r},0,t)]$.  From the general expression~(\ref{eq:gen-Ahydro}) a set of coupled equations for local nonequilibrium species densities and total number and momentum densities can be written which depend on their corresponding conjugate fields. The solutions of these equations can then be inserted into the expressions for the diffusiophoretic force and torque to complete the calculation of these quantities.

To illustrate how to carry out this program, consider the equation of motion for the average species number density fields $n_{\gamma}(\bm{r},t)$. For simplicity, we suppose the P\'{e}clet number is small, $Pe=V_{\rm sd}R_c/D_\gamma \ll 1$, so that advective effects can be neglected. Here $R_c$ is the colloid radius and $D_\gamma$ is the diffusion coefficient of species $\gamma$. When advective effects are small, the evolution of the number densities is independent of the fluid flow field.  In this case the evolution equation reads
\begin{eqnarray}\label{eq:gen-Ngamma}
&&\partial_t n_{\gamma}(\bm{r},t)= f_{\gamma,t}(\bm{r},t)  \\
&-& \int_0^t dt_1 \; \langle {\mathcal{F}}_{\gamma,t}(\bm{r},t_1,t) {\mathcal{F}}_{\gamma',t_1}(\bm{r}')
\rangle_{t_1} \ast \tilde{\mu}_{\gamma'}(\bm{r}',t_1),\nonumber
\end{eqnarray}
where we have used the summation convention and
\begin{eqnarray}\label{eq:random-force-gamma}
{\mathcal{F}}_{\gamma,t}(\bm{r},t_1,t)&=&U_{Q_A}(t_1,t) \mathcal{Q}_{A}(t)i \mathcal{L}_0 {N}_\gamma(\bm{r})\nonumber \\
&=&U_{Q_A}(t_1,t) {\mathcal{F}}_{\gamma,t}(\bm{r}),
\end{eqnarray}
with the random force given by $f_{\gamma,t}(\bm{r},t)=\Tr[\rho_b(0) {\mathcal{F}}_{\gamma,t}(\bm{r},0,t)]$. The random force vanishes if the initial condition is the local equilibrium distribution and will be neglected here.

The ${\mathcal{F}}_{\gamma,t}(\bm{r},t_1,t)$ functions evolve on a short time scale $\tau_m$ in view of the projected dynamics. Consequently, the time-ordered evolution operator $U_{Q_A}(t_1,t)$ can be simplified by replacing the projectors $\mathcal{Q}_{A}(t_n)$ by $\mathcal{Q}_{A}(t)$ so that $U_{Q_A}(t_1,t)\approx e^{\mathcal{Q}_{A}(t)i \mathcal{L}_0 (t-t_1)}$. Using this approximation and making the substitution $t_1=t-\tau$ in the integral we have
\begin{eqnarray}\label{eq:kernel-approx}
&&\int_0^t dt_1 \; \langle {\mathcal{F}}_{\gamma,t}(\bm{r},t_1,t) {\mathcal{F}}_{\gamma',t_1}(\bm{r}') \rangle_{t_1} \ast \tilde{\mu}_{\gamma'}(\bm{r}',t_1) \nonumber \\
&\approx& \Big[\int_0^\infty d\tau \; \langle \big(e^{\mathcal{Q}_{A}(t)i \mathcal{L}_0 \tau}{\mathcal{F}}_{\gamma,t}(\bm{r})\big) {\mathcal{F}}_{\gamma',t}(\bm{r}') \rangle_{t}\Big] \ast \tilde{\mu}_{\gamma'}(\bm{r}',t),\nonumber \\
\end{eqnarray}
where in the last line we replaced the upper limit $t$ in the integral by infinity for $t \gg \tau_m$, and replaced $\tilde{\mu}_\gamma(\bm{r}',t-\tau)$ by $\tilde{\mu}_{\gamma}(\bm{r}',t)$.

Using Eqs.~(\ref{eq:Ngamma-flux}) and (\ref{eq:rate-flux}) to obtain ${\mathcal{F}}_{\gamma,t}(\bm{r})=J_\gamma^R(\bm{r}) -\bm{\nabla{}}_r\cdot{}\bm{j}_{\gamma}(\bm{r})$, along with Eq.~(\ref{eq:kernel-approx}), and neglecting cross coupling between reaction and diffusion, the generalized reaction-diffusion equation~(\ref{eq:gen-Ngamma}) can be written as
\begin{eqnarray}\label{eq:RD-nonlocal}
\partial_t n_{\gamma}(\bm{r},t)&=& -L^R_{\gamma \gamma'}(\bm{r},\bm{r}')\ast \beta \tilde{\mu}_{\gamma'}(\bm{r}',t)\\
&&+\bm{\nabla_r}\cdot \bm{L}_{\gamma \gamma'}(\bm{r},\bm{r}')\ast \beta \bm{\nabla_{r'}}\tilde{\mu}_{\gamma'}(\bm{r}',t),\nonumber
\end{eqnarray}
where the reaction and diffusion transport terms are
\begin{eqnarray}
L^R_{\gamma \gamma'}(\bm{r},\bm{r}')&=&\int_0^\infty d\tau \; \langle J_\gamma^R(\bm{r},\tau^*) J_{\gamma'}^R(\bm{r}') \rangle_t,\\
\bm{L}_{\gamma \gamma'}(\bm{r},\bm{r}')&=&\int_0^\infty d\tau \; \langle \bm{j}_\gamma(\bm{r},\tau^*) \bm{j}_{\gamma'}(\bm{r}')\rangle_t,
\end{eqnarray}
and $\tau^*$ is used to denote evolution by projected dynamics. Since the chemical species are dilute in the solvent the constraint condition Eq.~(\ref{eq:gen-hydro-base}) relating the nonequilibrium species densities $n_\gamma (\bm{r},t)$ at time $t$ to the conjugate fields $\bm{\phi}(\bm{r},t)$  can be inverted to leading order in the fugacities of the dilute species.  As a result, the chemical potentials can be written as $\mu_\gamma(\bm{r},t)=\mu_\gamma^0 +k_BT\ln (n_{\gamma}(\bm{r},t)/n_0)$ and substitutions into Eq.~(\ref{eq:RD-nonlocal}) yields closed equations for these local species density fields in the presence of the fixed colloid. For our self-diffusiophoretic colloid these equations should be solved subject to constraints on the concentration fields at the boundaries where the system is in contact with reservoirs with fixed chemical concentrations. Although the solution of the fluid equations depends on where the colloid is located relative to the reservoirs, the behavior of the fluid densities in the vicinity of the colloid is determined by the local microscopic interactions of the fluid particles. An analogous treatment can be applied to the equations for the total number and momentum density fields. These transport equations, along with the expressions given above for the diffusiophoretic force and torque and colloid friction, provide a fully microscopic Langevin description of active self-diffusiophoretic dynamics.

For particles that are large compared to solvent species it is appropriate to describe interactions of the fluid species with the colloid through boundary conditions.~\cite{ALP82,A89,GK18b} As noted above, the behavior of the fluid densities in the vicinity of the colloid is determined by the local microscopic interactions of the fluid particles with the colloid.  These fluid densities typically exhibit rapid variations and structural and dynamic correlations near the colloid due to strong interactions of solvent particles at short distances from interaction sites on the colloid. The determination of the appropriate boundary conditions that account for the complicated surface structure and dynamics in the fluid induced by the colloidal interactions requires a detailed analysis of the generalized reaction-diffusion and hydrodynamic equations in the interaction zone around the colloid. Through such analyses the present microscopic description can be linked to continuum treatments of self-diffusiophoresis for large colloidal particles.

\section{Conclusion}\label{sec:conclusion}

The molecular-level derivation of the Langevin equations given in this paper for an active particle whose propulsion arises from a diffusiophoretic mechanism allows one to assess the domain of validity of Langevin descriptions of such active systems that are often proposed on phenomenological grounds. The generalized Langevin equations incorporate features that become important on small length and time scales.
These include the static structural correlations among fluid species and the active particle that complicate descriptions in which the effect of the colloid on the fluid is incorporated into boundary conditions, explicit treatment of both solvent and solute species and their interactions with the active particle, and memory effects that enter because the time scales of the dynamics are not as well separated as when the active particle is orders of magnitude larger than the solvent species.

The diffusiophoretic force and torque in the Langevin equations are important quantities that differentiate these Langevin equations from those for ordinary Brownian motion. They contain contributions that depend on the local nonequilibrium averages of species density fields expected from continuum calculations; however, these fields themselves satisfy generalized hydrodynamic and reaction-diffusion equations. In addition, they have contributions that involve full nonequilibrium averages of the reactive molecules that cannot be expressed separately in terms of the fuel and product species densities.

Another important feature that emerges from the microscopic derivation is that all transport and dynamical diffusiophoretic factors have microscopic expressions in terms of Green-Kubo correlation functions. This permits one, at least in principle, to determine these quantities directly from molecular dynamics simulations by numerically evaluating autocorrelation functions of the force and torque imparted on the fixed colloid by the fluid. Thus, the transport properties that enter in phenomenological Langevin models are specified in molecular terms. In particular, since the reactive species are treated at a molecular level that explicitly accounts for the dynamics of the nuclei comprising the molecules, the activated rate processes that take place on the colloid (or in the fluid) can be described in terms of suitable reaction coordinates whose specific forms depend on the reaction mechanism, and reaction rates can be computed using molecular dynamics employing rare event sampling methods for these slow processes.

It is simple to extend the formalism presented here to describe thermophoretically-active colloids in the presence of an external temperature gradient or to incorporate reactive events that are not iso-enthalpic.  In addition, while most of the presentation in this paper considered a rigid colloid, the development is not restricted to this specific kind of active particle. The active particle may be any molecule or molecular aggregate with internal degrees of freedom, so that the generalized Langevin equations presented can form a basis for the analysis of molecular simulations and experiments dealing with active diffusiophoretic motion on molecular scales.

\section*{Data Availability}
The data that support the findings of this study are available from the corresponding author upon reasonable request.

\section*{Acknowledgments}
Research was supported in part by grants from the Natural Sciences and Engineering Research Council of Canada. Financial support from the Universit\'e libre de Bruxelles (ULB) and the Fonds de la Recherche Scientifique~-~FNRS under the Grant PDR~T.0094.16 for the project ``SYMSTATPHYS" is also acknowledged.

\appendix

\section{Evolution operator}\label{sec:evolution-operator}

The full system evolution operator satisfies the equation
\begin{equation}
\partial_t U(0,t)= U(0,t) i\Liou{},
\end{equation}
while the time-ordered projected evolution operator $U_Q(0,t)$ satisfies
\begin{equation}
\partial_t U_Q(0,t)= U_Q(0,t) i\Liou{} \mathcal{Q}(t),
\end{equation}
whose formal solution can be written as
\begin{equation}
U_Q(0,t) = {\cal T}_{-} \exp \left( \int_0^t dt_1 \, i\Liou{} \mathcal{Q}(t_1) \right),
\end{equation}
where ${\cal T}_{-}$ is a time-ordering operator that orders operators in increasing order by their time argument.

The relation between these propagators can be established as follows: The evolution operators $U(0,t)$ and $U_Q(0,t)$ have the property $U(t_1,t_2)U(t_2,t_3)=U(t_1,t_3)$, with an analogous expression for $U_Q(t_1,t_2)$. To establish the relation between these operators we let $U(0,t)= G(t)U_Q(0,t)$ so that the operator $G(t)=U(0,t)U^{-1}_Q(0,t)$. Its initial value is $G(0)=1$. Since $U_Q(0,t)U^{-1}_Q(0,t)=1$ by definition, $U^{-1}_Q(0,t)$ satisfies
\begin{equation}
\partial_t U^{-1}_Q(0,t) = - i\Liou \mathcal{Q}(t) U^{-1}_Q(0,t).
\end{equation}
Differentiation of the definition of $G(t)$ yields
\begin{eqnarray}
\partial_t G(t)&=& U(0,t)i\Liou{}U^{-1}_Q(0,t) +U(0,t)\partial_t U^{-1}_Q(t,0)\nonumber \\
&=& U(0,t)i\Liou{} \mathcal{P}(t) U^{-1}_Q(0,t),
\end{eqnarray}
which, after integration, gives
\begin{eqnarray}
 G(t)&=& 1+ \int_0^t dt_1 \; U(0,t_1)i\Liou{} \mathcal{P}(t_1) U^{-1}_Q(0,t_1) \\
 &=& 1+ \int_0^t dt_1 \; (\partial_{t_1}U(0,t_1)) \mathcal{P}(t_1) U^{-1}_Q(0,t_1).\nonumber
\end{eqnarray}
Using this result we may then obtain $U(0,t)$ as
\begin{eqnarray}
U(0,t)&=&  U_Q(0,t) + \int_0^t dt_1 \; (\partial_{t_1}U(0,t_1)) \mathcal{P}(t_1) U_Q(t_1,t)\nonumber \\
&=&U_Q(0,t) + \int_0^t dt_1 \; \partial_{t_1}(U(0,t_1) \mathcal{P}(t_1) U_Q(t_1,t))\nonumber\\
&&- \int_0^t dt_1 \; U(0,t_1) \partial_{t_1}(\mathcal{P}(t_1) U_Q(t_1,t)),
\end{eqnarray}
which can be rearranged to give Eq.~(\ref{eq:U-UQ-relation}) in the main text.

\section{Reduction of integral terms in Eq.~(\ref{eq:PLang1})}\label{sec:integral-terms}
Let ${\it I}$ denote the integral terms in Eq.~(\ref{eq:PLang1}),
\begin{equation}\label{eq:intred0}
{\it I}=\int_0^t dt_1 \; e^{i\Liou{} t_1} [\mathcal{P}(t_1) i\Liou{} \bm{K}_D(t_1,t) -(\partial_{t_1}\mathcal{P}(t_1)) \bm{K}_D(t_1,t)].
\end{equation}
We have
\begin{eqnarray}
&&-(\partial_{t_1}\mathcal{P}(t_1)) \bm{K}_D(t_1,t)=\Tr[(i \Liou{}_0 \rho_b(t_1)) \bm{K}_D(t_1,t)]\nonumber\\
&& \qquad \qquad
=-\Tr[\rho_b(t_1) i \Liou{}_0 \bm{K}_D(t_1,t)] \nonumber \\
&&\qquad \qquad =-\mathcal{P}(t_1) i\Liou{}_0 \bm{K}_D(t_1,t).
\end{eqnarray}
Using this result, along with $i\Liou{} -i\Liou{}_0 =i\Liou{}_c$,  Eq.~(\ref{eq:intred0}) can be written as
\begin{eqnarray} \label{eq:intred2}
{\it I} &=& \int_0^t dt_1 \; e^{i\Liou{} t_1} \mathcal{P}(t_1) i\Liou{}_c \bm{K_D}(t_1,t) \nonumber \\
&=& \int_0^t dt_1 \; e^{i\Liou{} t_1} \Tr [\rho_b(t_1)\Big(
\frac{\bm{P}}{M} \cdot \bm{\nabla}_R \nonumber \\
&& \qquad \qquad +\bm{F}_c \cdot \bm{\nabla}_P + i{\cal L}_{\rm rot}\Big)\bm{K}_D(t_1,t).
\end{eqnarray}
Also,
\begin{eqnarray}
&&\Tr \left[ \rho_b(t_1)\frac{\bm{P}}{M} \cdot \bm{\nabla}_R \bm{K}_D (t_1,t) \right]= \nonumber \\
&& \qquad \qquad -\frac{\bm{P}}{M} \cdot \Tr \left[ (\bm{\nabla}_R \rho_b(t_1))
\bm{K}_D(t_1,t)\right] \nonumber \\
&& \Tr \left[ \rho_b(t_1) i{\cal L}_{rot} \bm{K}_{D}(t_1,t) \right] = \nonumber \\
&&\qquad  \qquad -\bm{\Pi}^T \cdot \bm{M}^{-1} \cdot \Tr \left[ \bm{\nabla}_\theta \rho_b(t_1)
\bm{K}_D (t_1,t) \right] \nonumber \\
&& \qquad \qquad - \bm{\nabla}_{\Pi} \cdot \Tr \left[
\rho_b(t_1) \bm{\nabla}_\theta U_I \bm{K}_D(t_1,t) \right],
\end{eqnarray}
since $\Tr [\rho_b(t_1)  \bm{K}_D(t_1,t)]=0$. We then have
\begin{eqnarray}\label{eq:intred4}
&&{\it I} =\int_0^t dt_1 \; e^{i\Liou{} t_1} \Big\{
 -\frac{\bm{P}}{M} \cdot \bm{M}_1(t_1,t) + \bm{\nabla}_P \cdot \bm{M}_2(t_1,t) \nonumber\\
 &-&\bm{\Pi}^T \cdot \bm{M}^{-1} \cdot M_3(t_1,t) -
 \bm{\nabla}_{\Pi} \cdot M_4(t_1,t)\Big\},\nonumber \\
&=& \int_0^t dt_1 \;  \Big\{ -\frac{\bm{P}(t_1)}{M} \cdot \bm{M}_1(t_1,t ))  + \bm{\nabla}_P(t_1) \cdot \bm{M}_2(t_1,t) \nonumber\\
&-& \bm{\Pi}(t_1)^T \cdot \bm{M}(t_1)^{-1} \cdot \bm{M}_3(t_1,t) - \bm{\nabla}_{\Pi}(t_1) \cdot \bm{M}_4(t_1,t) \Big\},\nonumber \\
\end{eqnarray}
where $\bm{M}_1, \dots, \bm{M}_4$ are defined in Eq.~(\ref{eq:M-quantities}). Use of this expression yields Eq.~(\ref{eq:PLang1sup}).

\section{$\rho_b(t)$ and $\rho_L(t)$ densities}\label{sec:bathDensity}

An explicit relation between $\rho_b(t)$ and $\rho_L(t)$ is required in order to express average values in a convenient form. For this purpose we consider a projection operator
\begin{equation}\label{eq:b_proj-comp}
\mathcal{P}_{A}(t)f = \langle f\bm{C}(\bm{r}_{1})\rangle_t\ast{}\langle{}\bm{C}\bm{C}\rangle_{t}^{-1}(\bm{r}_{1},\bm{r}_{2})
\ast{}\bm{C}(\bm{r}_{2}) ,
\end{equation}
and its complement $\mathcal{Q}_{A}(t)=1-\mathcal{P}_{A}(t)$. The adjoint of this projector is  defined by
\begin{equation}\label{eq:le_proj}
\mathcal{P}^\dagger_{A}(t)f = \Tr[f\bm{C}(\bm{r}_{1})]\ast{}\langle{}\bm{C}\bm{C}\rangle_{t}^{-1}(\bm{r}_{1},\bm{r}_{2})\ast{}\bm{C}(\bm{r}_{2}) \rho_L(t).
\end{equation}
and its complement $\mathcal{Q}^\dagger_{A}(t)=1-\mathcal{P}^\dagger_{A}(t)$. The vector $\bm{C}(\bm{r})=\{1,\tilde{\bm{A}}(\bm{r})\}$ is expressed in terms of the deviations $\tilde{\bm{A}}(\bm{r})\equiv {\bm{A}}(\bm{r})-\langle{\bm{A}}(\bm{r})\rangle_t$ of the fields ${\bm{A}}(\bm{r})$ in Eq.~(\ref{eq:A-fields}). Using this notation, we observe that $1$ is not a field extending over space but a single number, while $\tilde{\bm{A}}(\bm{r})$ is a field variable; hence, we can write
\begin{eqnarray}\label{eq:le_proj-2}
\mathcal{P}^\dagger_{A}(t)f &=& \Tr[f]\rho_L(t)\\
&+& \Tr[f \tilde{\bm{A}}(\bm{r}_{1})]\ast{}\langle{}\tilde{\bm{A}}\tilde{\bm{A}}\rangle_{t}^{-1}(\bm{r}_{1},\bm{r}_{2})\ast{}\tilde{\bm{A}}(\bm{r}_{2}) \rho_L(t).\nonumber
\end{eqnarray}
Taking $f=\rho_b(t)$ and using the fact that $\Tr[ \rho_b(t) \bm{A}(\bm{r})]= \langle \bm{A}(\bm{r}) \rangle_t$ we obtain $\mathcal{P}^\dagger_{A}(t)\rho_b(t) =  \rho_L(t)$. We may then write $\rho_b(t)= \rho_L(t)+ \mathcal{Q}^\dagger_{A}(t) \rho_b(t)$.

Applying this projector to Eq.~(\ref{eq:liou-bath}), we have
\begin{eqnarray}
\mathcal{P}^\dagger_{A}(t) \partial_t \rho_b(t)&=&\Tr[(\partial_t \rho_b(t))\bm{C}(\bm{r}_{1})]\ast{}\langle{}\bm{C}\bm{C}\rangle_{t}^{-1}(\bm{r}_{1},\bm{r}_{2})\nonumber \\
&& \qquad \qquad \qquad \ast{}\bm{C}(\bm{r}_{2}) \rho_L(t)\nonumber \\
 &=& \Tr[(\partial_t \rho_L(t))\bm{C}(\bm{r}_{1})]\ast{}\langle{}\bm{C}\bm{C}\rangle_{t}^{-1}(\bm{r}_{1},\bm{r}_{2})\nonumber \\ && \qquad \qquad \qquad \ast{}\bm{C}(\bm{r}_{2}) \rho_L(t).
\end{eqnarray}

Since the local nonequilibrium distribution function may be written as
\begin{equation}\label{eq:local_eq2}
\rho_{L}(t)=\frac{\prod_\lambda (N_\lambda ! h^{3N_\lambda})^{-1}e^{\bm{C}(\bm{r}) \ast{} \bm{\phi}_{C}(\bm{r},t)}}{\Tr[
\prod_\lambda (N_\lambda ! h^{3N_\lambda})^{-1}e^{\bm{C}(\bm{r})\ast \bm{\phi}_{C}(\bm{r},t)}]},
\end{equation}
with $\bm{\phi_{C}}= (0,\bm{\phi_{A}})$, we have
\begin{eqnarray}
\partial_t \rho_L(t)&=& (\partial_t \bm{\phi_{C}}(\bm{r},t))\ast \bm{C}(\bm{r}) \rho_L(t)\nonumber \\
&=& (\partial_t \bm{\phi_{A}}(\bm{r},t))\ast \tilde{\bm{A}}(\bm{r}) \rho_L(t),
\end{eqnarray}
and
\begin{equation}
\mathcal{P}^\dagger_{A}(t) \partial_t \rho_b(t)=(\partial_t \bm{\phi_{C}}(\bm{r},t))\ast \bm{C}(\bm{r}) \rho_L(t)=\partial_t \rho_L(t).
\end{equation}
From this result we can write
\begin{eqnarray}\label{eq:rhob-evol}
\partial_t \rho_b(t)&=&-i \Liou{}_0 (\mathcal{P}^\dagger_{A}(t) \rho_b(t) + \mathcal{Q}^\dagger_{A}(t) \rho_b(t))\nonumber \\
&=&
-i \Liou{}_0 \rho_L(t) -i \Liou{}_0 \mathcal{Q}^\dagger_{A}(t) \rho_b(t),
\end{eqnarray}
and
\begin{equation}\label{eq:Qrhob-rel}
\mathcal{Q}^\dagger_{A}(t) \partial_t \rho_b(t)= \partial_t \rho_b(t)- \partial_t \rho_L(t) =\partial_t \mathcal{Q}^\dagger_{A}(t) \rho_b(t).
\end{equation}
Using Eqs.~(\ref{eq:rhob-evol}) and (\ref{eq:Qrhob-rel}) we have
\begin{equation}\label{eq:Qrhob-evol}
\partial_t \mathcal{Q}^\dagger_{A}(t) \rho_b(t)= - \mathcal{Q}^\dagger_{A}(t) i\Liou{}_0 \rho_L(t) - \mathcal{Q}^\dagger_{A}(t) i\Liou{}_0 \mathcal{Q}^\dagger_{A}(t) \rho_b(t).
\end{equation}

To solve this equation we introduce the projected propagator $U^\dagger_{{\cal Q}_A}(0,t)$  that is the solution of the evolution equation
\begin{equation}
\partial_t U^\dagger_{{\cal Q}_A}(0,t) = -\mathcal{Q}^\dagger_{A}(t) i \Liou_0 U^\dagger_{{\cal Q}_A}(0,t),
\label{eq:projectionEvolution}
\end{equation}
and its inverse $(U^\dagger_{{\cal Q}_A})^{-1}(0,t)$ whose evolution is given by
\begin{equation}
\partial_t (U^\dagger_{{\cal Q}_A})^{-1}(0,t) = (U^\dagger_{{\cal Q}_A})^{-1}(0,t) \mathcal{Q}^\dagger_{A}(t) i \Liou_0 U^\dagger_{{\cal Q}_A}(0,t) .
\end{equation}

Formally, the solution of Eq.~(\ref{eq:projectionEvolution}) can be written as
\begin{equation}
U^\dagger_{{\cal Q}_A}(0,t) = {\cal T}_+ \exp \left( -\int_0^t dt_1 \,
\mathcal{Q}^\dagger_{A}(t_1) i \Liou_0 \right) ,
\label{eq:formalUdagger}
\end{equation}
where the time-ordering operator ${\cal T}_+$ orders operators with smaller time argument to the right of operators of larger time argument.

Defining an operator $G(t)= (U^\dagger_{{\cal Q}_A})^{-1}(0,t) \mathcal{Q}^\dagger_{A}(t) \rho_b(t)$ and using the above results its differential equation is given by
\begin{equation}
\partial_t G(t) = -(U^\dagger_{{\cal Q}_A})^{-1}(0,t) \mathcal{Q}^\dagger_{A}(t) i \Liou_0 \rho_L(t).
\end{equation}
Integration of this equation gives
\begin{eqnarray}
G(t) &=& {\cal Q}_A^{\dagger}(0)\rho_b(0)  \\
&&- \int_0^t dt_1 \; (U^\dagger_{{\cal Q}_A})^{-1}(0,t_1)
\mathcal{Q}^\dagger_{A}(t_1) i\Liou{}_0 \rho_L(t_1),\nonumber
\end{eqnarray}
from which we find
\begin{eqnarray}
\mathcal{Q}^\dagger_{A}(t) \rho_b(t) &=& U^\dagger_{{\cal Q}_A}(t,0) \mathcal{Q}^\dagger_{A}(0) \rho_b(0) \\
 &&- \int_0^t dt_1 \; U^\dagger_{{\cal Q}_A}(t,t_1)
\mathcal{Q}^\dagger_{A}(t_1) i\Liou{}_0 \rho_L(t_1).\nonumber
\end{eqnarray}

It follows that
\begin{eqnarray}\label{eq:rhob-rhoL1}
\rho_b(t) &=& \rho_L(t) +U^\dagger_{{\cal Q}_A}(t,0) \mathcal{Q}^\dagger_{A}(0) \rho_b(0)  \\
&&- \int_0^t dt_1 \; U^\dagger_{{\cal Q}_A}(t,t_1)
\mathcal{Q}^\dagger_{A}(t_1) i\Liou{}_0 \rho_L(t_1).\nonumber
\end{eqnarray}
This equation may be written in another form by using $-i\Liou{}_0 \rho_L(t)=-\dot{\bm{A}}(\bm{r})\ast{}\bm{\phi_{A}}(\bm{r},t) \rho_L(t)$ and the fact that
$\mathcal{Q}_A^\dagger(t)i\Liou_0 {N}(\bm{r})\rho_L(t)=0$ along with $\mathcal{Q}_A^\dagger(t)i\Liou_0 \Big(\int d\bm{r}\;E_{N}(\bm{r})\Big)\rho_L(t)=0$ since $\phi_E(\bm{r},t)=-\beta$. We have
\begin{eqnarray}\label{eq:rhob-rhoL1bis}
&&\rho_b(t) = \rho_L(t) +U^\dagger_{{\cal Q}_A}(t,0) \mathcal{Q}^\dagger_{A}(0) \rho_b(0) - \\
&& \int_0^t dt_1 \; U^\dagger_{{\cal Q}_A}(t,t_1)\Big(
\mathcal{Q}_{A}(t_1) \dot{N}_\gamma(\bm{r})+\mathcal{Q}_{A}(t_1) \dot{\bm{g}}_N(\bm{r})\Big) \rho_L(t_1),\nonumber
\end{eqnarray}
where use of the identity $\mathcal{Q}^\dagger_{A}(t) i\Liou{}_0 f\rho_L(t)=\big(\mathcal{Q}_{A}(t_1) i\Liou{}_0 f \big)\rho_L(t_1)$ for some function $f$ has been made.

The fluxes in Eq.~(\ref{eq:rhob-rhoL1bis}) are $\dot{N}_{\gamma}(\bm{r})$ in Eq.~(\ref{eq:Ngamma-flux}) and
\begin{equation}\label{eq:momenuym-flux}
\dot{\bm{g}}_N(\bm{r}) =-\bm{\nabla{}}_r\cdot{}\bm{\tau{}}(\bm{r})+\bm{F}_{\rm f}(\bm{r}),
\end{equation}
where the fluid stress tensor is
\begin{eqnarray}\label{eq:stress-tensor}
&&\bm{\tau}(\bm{r})= \sum_{i=1}^N\Big[ \frac{\bm{p}_i\bm{p}_i}{m}-\frac{1}{2} \sum_{j\ne i}^N \Big(\sum_{\nu}\Theta_i^S \Theta_j^{\nu} \bm{r}_{ij} \bm{\nabla_r}V_{S\nu} \nonumber \\
&+&\Theta_i^{R} \Theta_j^{R} \bm{r}_{ij} \sum_{k,k'=1}^{n_a}\bm{\nabla}_{\bm{r}_i^{(k)}}V_{kk'}(|\bm{r}_i^{(k)}-\bm{r}_j^{(k')}|)\Big)
\Big]\delta(\bm{r}_{ic}-\bm{r}) \nonumber\\
\end{eqnarray}
in the small gradient approximation~\cite{M67}, and the local force on the fluid is
\begin{eqnarray} \label{eq:force-on-fluid}
\bm{F}_{\rm f}(\bm{r})&=& -\sum_{i=1}^N\Big[\sum_{b=N}^C\sum_{\alpha =1}^{n_s} \Theta_\alpha^b \Big(\Theta_i^S \bm{\nabla_r} V_{Sb}(r^\alpha)\\
&&+\Theta_i^R \sum_{k=1}^{n_a}\bm{\nabla}_{\bm{r}_i^{ak}}V_{kb}({r}_{ic}^{(k)\alpha})\Big)
\Big]\delta(\bm{r}_{ic}-\bm{r}) . \nonumber
\end{eqnarray}
We may then write Eq.~(\ref{eq:rhob-rhoL1bis}) for an isothermal system as
\begin{eqnarray}\label{eq:rhob-rhoL2}
&&\rho_b(t) = \rho_L(t) +U^\dagger_{Q_A}(t,0) \mathcal{Q}^\dagger_{A}(0) \rho_b(0)  \\
&-&\beta \int_0^t dt_1 \; U^\dagger_{Q_A}(t,t_1)
\Big(\mathcal{Q}_A(t_1)J_\gamma^R(\bm{r})\ast \tilde{\mu{}}_{\gamma{}}(\bm{r},t_1)\nonumber \\
&&\quad + \mathcal{Q}_A(t_1)\bm{j}_{\gamma}(\bm{r})\ast  \bm{\nabla}_r \tilde{\mu{}}_{\gamma{}}(\bm{r},t_1) \nonumber \\
&&\quad + \mathcal{Q}_A(t_1)\bm{\tau{}}(\bm{r}) \ast  \bm{\nabla{}}_r \bm{v}(\bm{r},t_1)  \nonumber\\
&&\quad + \mathcal{Q}_A(t_1)\bm{F}_{\rm f}(\bm{r}) \ast  \bm{v}(\bm{r},t_1)  \Big) \rho_L(t_1),\nonumber
\end{eqnarray}
which is the relation we sought. The initial condition term will decay on a molecular time scale in view of the projected evolution. Also, if the initial condition is $\rho_b(0)=\rho_L(0)$ this term is identically zero. Thus we can neglect it in the computation. Also, Since $J^R(\bm{r}) \equiv J_A^R(\bm{r})=-J_B^R(\bm{r})$ we can express the reactive contribution in terms of the chemical affinity $\mathcal{A}(\bm{r},t)=\mu_A (\bm{r},t)-\mu_B (\bm{r},t)$ as
\begin{equation}
\mathcal{Q}_A(t_1)J_\gamma^R(\bm{r})\ast \tilde{\mu{}}_{\gamma{}}(\bm{r},t_1)=\mathcal{Q}_A(t_1)J^R(\bm{r})\ast \mathcal{A}(\bm{r},t).
\end{equation}

\bibliography{langevin-refs}

\begin{thebibliography}{54}%
\makeatletter
\providecommand \@ifxundefined [1]{%
 \@ifx{#1\undefined}
}%
\providecommand \@ifnum [1]{%
 \ifnum #1\expandafter \@firstoftwo
 \else \expandafter \@secondoftwo
 \fi
}%
\providecommand \@ifx [1]{%
 \ifx #1\expandafter \@firstoftwo
 \else \expandafter \@secondoftwo
 \fi
}%
\providecommand \natexlab [1]{#1}%
\providecommand \enquote  [1]{``#1''}%
\providecommand \bibnamefont  [1]{#1}%
\providecommand \bibfnamefont [1]{#1}%
\providecommand \citenamefont [1]{#1}%
\providecommand \href@noop [0]{\@secondoftwo}%
\providecommand \href [0]{\begingroup \@sanitize@url \@href}%
\providecommand \@href[1]{\@@startlink{#1}\@@href}%
\providecommand \@@href[1]{\endgroup#1\@@endlink}%
\providecommand \@sanitize@url [0]{\catcode `\\12\catcode `\$12\catcode
  `\&12\catcode `\#12\catcode `\^12\catcode `\_12\catcode `\%12\relax}%
\providecommand \@@startlink[1]{}%
\providecommand \@@endlink[0]{}%
\providecommand \url  [0]{\begingroup\@sanitize@url \@url }%
\providecommand \@url [1]{\endgroup\@href {#1}{\urlprefix }}%
\providecommand \urlprefix  [0]{URL }%
\providecommand \Eprint [0]{\href }%
\providecommand \doibase [0]{http://dx.doi.org/}%
\providecommand \selectlanguage [0]{\@gobble}%
\providecommand \bibinfo  [0]{\@secondoftwo}%
\providecommand \bibfield  [0]{\@secondoftwo}%
\providecommand \translation [1]{[#1]}%
\providecommand \BibitemOpen [0]{}%
\providecommand \bibitemStop [0]{}%
\providecommand \bibitemNoStop [0]{.\EOS\space}%
\providecommand \EOS [0]{\spacefactor3000\relax}%
\providecommand \BibitemShut  [1]{\csname bibitem#1\endcsname}%
\let\auto@bib@innerbib\@empty
\bibitem [{\citenamefont {Ramaswamy}(2010)}]{R10}%
  \BibitemOpen
  \bibfield  {author} {\bibinfo {author} {\bibfnamefont {S.}~\bibnamefont
  {Ramaswamy}},\ }\href@noop {} {\bibfield  {journal} {\bibinfo  {journal}
  {Annu. Rev. Condens. Matter Phys.}\ }\textbf {\bibinfo {volume} {1}},\
  \bibinfo {pages} {323} (\bibinfo {year} {2010})}\BibitemShut {NoStop}%
\bibitem [{\citenamefont {Vicsek}\ and\ \citenamefont {Zafeiris}(2012)}]{V12}%
  \BibitemOpen
  \bibfield  {author} {\bibinfo {author} {\bibfnamefont {T.}~\bibnamefont
  {Vicsek}}\ and\ \bibinfo {author} {\bibfnamefont {A.}~\bibnamefont
  {Zafeiris}},\ }\href {\doibase 10.1016/j.physrep.2012.03.004} {\bibfield
  {journal} {\bibinfo  {journal} {Phys. Rep.}\ }\textbf {\bibinfo {volume}
  {517}},\ \bibinfo {pages} {71} (\bibinfo {year} {2012})}\BibitemShut
  {NoStop}%
\bibitem [{\citenamefont {Aronson}(2013)}]{A13}%
  \BibitemOpen
  \bibfield  {author} {\bibinfo {author} {\bibfnamefont {I.~S.}\ \bibnamefont
  {Aronson}},\ }\href {\doibase 10.1016/j.crhy.2013.05.002} {\bibfield
  {journal} {\bibinfo  {journal} {Comptes Rendus Physique}\ }\textbf {\bibinfo
  {volume} {14}},\ \bibinfo {pages} {518} (\bibinfo {year} {2013})}\BibitemShut
  {NoStop}%
\bibitem [{\citenamefont {Elgeti}, \citenamefont {Winkler},\ and\ \citenamefont
  {Gompper}(2015)}]{EWG15}%
  \BibitemOpen
  \bibfield  {author} {\bibinfo {author} {\bibfnamefont {J.}~\bibnamefont
  {Elgeti}}, \bibinfo {author} {\bibfnamefont {R.~G.}\ \bibnamefont {Winkler}},
  \ and\ \bibinfo {author} {\bibfnamefont {G.}~\bibnamefont {Gompper}},\ }\href
  {\doibase 10.1088/0034-4885/78/5/056601} {\bibfield  {journal} {\bibinfo
  {journal} {Rep. Prog. Phys.}\ }\textbf {\bibinfo {volume} {78}},\ \bibinfo
  {pages} {056601} (\bibinfo {year} {2015})}\BibitemShut {NoStop}%
\bibitem [{\citenamefont {Fodor}\ and\ \citenamefont {Marchetti}(2018)}]{FM18}%
  \BibitemOpen
  \bibfield  {author} {\bibinfo {author} {\bibfnamefont {E.}~\bibnamefont
  {Fodor}}\ and\ \bibinfo {author} {\bibfnamefont {M.~C.}\ \bibnamefont
  {Marchetti}},\ }\href@noop {} {\bibfield  {journal} {\bibinfo  {journal}
  {Physica A}\ }\textbf {\bibinfo {volume} {504}},\ \bibinfo {pages} {106}
  (\bibinfo {year} {2018})}\BibitemShut {NoStop}%
\bibitem [{\citenamefont {Z\"{o}ttl}\ and\ \citenamefont {Stark}(2016)}]{ZS16}%
  \BibitemOpen
  \bibfield  {author} {\bibinfo {author} {\bibfnamefont {A.}~\bibnamefont
  {Z\"{o}ttl}}\ and\ \bibinfo {author} {\bibfnamefont {H.}~\bibnamefont
  {Stark}},\ }\href {\doibase 10.1088/0953-8984/28/25/253001} {\bibfield
  {journal} {\bibinfo  {journal} {J. Phys.: Condens. Matter}\ }\textbf
  {\bibinfo {volume} {28}},\ \bibinfo {pages} {253001} (\bibinfo {year}
  {2016})}\BibitemShut {NoStop}%
\bibitem [{\citenamefont {Dukhin}\ and\ \citenamefont
  {Derjaguin}(1974)}]{DD74}%
  \BibitemOpen
  \bibfield  {author} {\bibinfo {author} {\bibfnamefont {S.~S.}\ \bibnamefont
  {Dukhin}}\ and\ \bibinfo {author} {\bibfnamefont {B.~V.}\ \bibnamefont
  {Derjaguin}},\ }\href@noop {} {\emph {\bibinfo {title} {Surface and Colloid
  Science, Edited by {E.~Matijevic}, Vol. 7, Chap. 3}}}\ (\bibinfo  {publisher}
  {Wiley},\ \bibinfo {address} {New York},\ \bibinfo {year} {1974})\BibitemShut
  {NoStop}%
\bibitem [{\citenamefont {Anderson}(1986)}]{A86}%
  \BibitemOpen
  \bibfield  {author} {\bibinfo {author} {\bibfnamefont {J.~L.}\ \bibnamefont
  {Anderson}},\ }\href@noop {} {\bibfield  {journal} {\bibinfo  {journal} {Ann.
  N. Y. Acad. Sci.}\ }\textbf {\bibinfo {volume} {469}},\ \bibinfo {pages}
  {166} (\bibinfo {year} {1986})}\BibitemShut {NoStop}%
\bibitem [{\citenamefont {Anderson}(1989)}]{A89}%
  \BibitemOpen
  \bibfield  {author} {\bibinfo {author} {\bibfnamefont {J.~L.}\ \bibnamefont
  {Anderson}},\ }\href@noop {} {\bibfield  {journal} {\bibinfo  {journal} {Ann.
  Rev. Fluid Mech.}\ }\textbf {\bibinfo {volume} {21}},\ \bibinfo {pages} {61}
  (\bibinfo {year} {1989})}\BibitemShut {NoStop}%
\bibitem [{\citenamefont {Anderson}, \citenamefont {Lowell},\ and\
  \citenamefont {Prieve}(1982)}]{ALP82}%
  \BibitemOpen
  \bibfield  {author} {\bibinfo {author} {\bibfnamefont {J.~L.}\ \bibnamefont
  {Anderson}}, \bibinfo {author} {\bibfnamefont {M.~E.}\ \bibnamefont
  {Lowell}}, \ and\ \bibinfo {author} {\bibfnamefont {D.~C.}\ \bibnamefont
  {Prieve}},\ }\href@noop {} {\bibfield  {journal} {\bibinfo  {journal} {J.
  Fluid Mech.}\ }\textbf {\bibinfo {volume} {117}},\ \bibinfo {pages} {107}
  (\bibinfo {year} {1982})}\BibitemShut {NoStop}%
\bibitem [{\citenamefont {Golestanian}, \citenamefont {Liverpool},\ and\
  \citenamefont {Ajdari}(2005)}]{GLA05}%
  \BibitemOpen
  \bibfield  {author} {\bibinfo {author} {\bibfnamefont {R.}~\bibnamefont
  {Golestanian}}, \bibinfo {author} {\bibfnamefont {T.~B.}\ \bibnamefont
  {Liverpool}}, \ and\ \bibinfo {author} {\bibfnamefont {A.}~\bibnamefont
  {Ajdari}},\ }\href@noop {} {\bibfield  {journal} {\bibinfo  {journal} {Phys.
  Rev. Lett.}\ }\textbf {\bibinfo {volume} {94}},\ \bibinfo {pages} {220801}
  (\bibinfo {year} {2005})}\BibitemShut {NoStop}%
\bibitem [{\citenamefont {Kapral}(2013)}]{K13}%
  \BibitemOpen
  \bibfield  {author} {\bibinfo {author} {\bibfnamefont {R.}~\bibnamefont
  {Kapral}},\ }\href@noop {} {\bibfield  {journal} {\bibinfo  {journal} {J.
  Chem. Phys.}\ }\textbf {\bibinfo {volume} {138}},\ \bibinfo {pages} {020901}
  (\bibinfo {year} {2013})}\BibitemShut {NoStop}%
\bibitem [{\citenamefont {Colberg}\ \emph {et~al.}(2014)\citenamefont
  {Colberg}, \citenamefont {Reigh}, \citenamefont {Robertson},\ and\
  \citenamefont {Kapral}}]{CRRK14}%
  \BibitemOpen
  \bibfield  {author} {\bibinfo {author} {\bibfnamefont {P.~H.}\ \bibnamefont
  {Colberg}}, \bibinfo {author} {\bibfnamefont {S.~Y.}\ \bibnamefont {Reigh}},
  \bibinfo {author} {\bibfnamefont {B.}~\bibnamefont {Robertson}}, \ and\
  \bibinfo {author} {\bibfnamefont {R.}~\bibnamefont {Kapral}},\ }\href@noop {}
  {\bibfield  {journal} {\bibinfo  {journal} {Acc. Chem. Res.}\ }\textbf
  {\bibinfo {volume} {47}},\ \bibinfo {pages} {3504} (\bibinfo {year}
  {2014})}\BibitemShut {NoStop}%
\bibitem [{\citenamefont {Popescu}, \citenamefont {Uspal},\ and\ \citenamefont
  {Dietrich}(2016)}]{PUD16}%
  \BibitemOpen
  \bibfield  {author} {\bibinfo {author} {\bibfnamefont {M.~N.}\ \bibnamefont
  {Popescu}}, \bibinfo {author} {\bibfnamefont {W.~E.}\ \bibnamefont {Uspal}},
  \ and\ \bibinfo {author} {\bibfnamefont {S.}~\bibnamefont {Dietrich}},\
  }\href {\doibase 10.1140/epjst/e2016-60058-2} {\bibfield  {journal} {\bibinfo
   {journal} {The European Physical Journal Special Topics}\ }\textbf {\bibinfo
  {volume} {225}},\ \bibinfo {pages} {2189} (\bibinfo {year}
  {2016})}\BibitemShut {NoStop}%
\bibitem [{\citenamefont {Bechinger}\ \emph {et~al.}(2016)\citenamefont
  {Bechinger}, \citenamefont {Leonardo}, \citenamefont {L\"{o}wen},
  \citenamefont {Reichhardt}, \citenamefont {Volpe},\ and\ \citenamefont
  {Volpe}}]{BDLRVV16}%
  \BibitemOpen
  \bibfield  {author} {\bibinfo {author} {\bibfnamefont {C.}~\bibnamefont
  {Bechinger}}, \bibinfo {author} {\bibfnamefont {R.~D.}\ \bibnamefont
  {Leonardo}}, \bibinfo {author} {\bibfnamefont {H.}~\bibnamefont {L\"{o}wen}},
  \bibinfo {author} {\bibfnamefont {C.}~\bibnamefont {Reichhardt}}, \bibinfo
  {author} {\bibfnamefont {G.}~\bibnamefont {Volpe}}, \ and\ \bibinfo {author}
  {\bibfnamefont {G.}~\bibnamefont {Volpe}},\ }\href {\doibase
  10.1103/revmodphys.88.045006} {\bibfield  {journal} {\bibinfo  {journal}
  {Rev. Mod. Phys.}\ }\textbf {\bibinfo {volume} {88}},\ \bibinfo {pages}
  {045006} (\bibinfo {year} {2016})}\BibitemShut {NoStop}%
\bibitem [{\citenamefont {Stark}(2019)}]{S19}%
  \BibitemOpen
  \bibfield  {author} {\bibinfo {author} {\bibfnamefont {H.}~\bibnamefont
  {Stark}},\ }\href {\doibase 10.1021/acs.accounts.8b002599} {\bibfield
  {journal} {\bibinfo  {journal} {Acc. Chem. Res.}\ }\textbf {\bibinfo {volume}
  {51}},\ \bibinfo {pages} {2681} (\bibinfo {year} {2019})}\BibitemShut
  {NoStop}%
\bibitem [{\citenamefont {Wang}(2013)}]{W13}%
  \BibitemOpen
  \bibfield  {author} {\bibinfo {author} {\bibfnamefont {J.}~\bibnamefont
  {Wang}},\ }\href@noop {} {\emph {\bibinfo {title} {Nanomachines: Fundamentals
  and Applications}}}\ (\bibinfo  {publisher} {Wiley-VCH},\ \bibinfo {address}
  {Weinheim, Germany},\ \bibinfo {year} {2013})\BibitemShut {NoStop}%
\bibitem [{\citenamefont {Wang}\ \emph {et~al.}(2013)\citenamefont {Wang},
  \citenamefont {Duan}, \citenamefont {Ahmed}, \citenamefont {Mallouk},\ and\
  \citenamefont {Sen}}]{WDAMS13}%
  \BibitemOpen
  \bibfield  {author} {\bibinfo {author} {\bibfnamefont {W.}~\bibnamefont
  {Wang}}, \bibinfo {author} {\bibfnamefont {W.}~\bibnamefont {Duan}}, \bibinfo
  {author} {\bibfnamefont {S.}~\bibnamefont {Ahmed}}, \bibinfo {author}
  {\bibfnamefont {T.~E.}\ \bibnamefont {Mallouk}}, \ and\ \bibinfo {author}
  {\bibfnamefont {A.}~\bibnamefont {Sen}},\ }\href {\doibase
  10.1016/j.nantod.2013.08.009} {\bibfield  {journal} {\bibinfo  {journal}
  {Nano Today}\ }\textbf {\bibinfo {volume} {8}},\ \bibinfo {pages} {531}
  (\bibinfo {year} {2013})}\BibitemShut {NoStop}%
\bibitem [{\citenamefont {S\'anchez}, \citenamefont {Soler},\ and\
  \citenamefont {Katuri}(2015)}]{SSK15}%
  \BibitemOpen
  \bibfield  {author} {\bibinfo {author} {\bibfnamefont {S.}~\bibnamefont
  {S\'anchez}}, \bibinfo {author} {\bibfnamefont {L.}~\bibnamefont {Soler}}, \
  and\ \bibinfo {author} {\bibfnamefont {J.}~\bibnamefont {Katuri}},\
  }\href@noop {} {\bibfield  {journal} {\bibinfo  {journal} {Angew. Chem. Int.
  Ed.}\ }\textbf {\bibinfo {volume} {54}},\ \bibinfo {pages} {1414} (\bibinfo
  {year} {2015})}\BibitemShut {NoStop}%
\bibitem [{\citenamefont {Wong}, \citenamefont {Dey},\ and\ \citenamefont
  {Sen}(2016)}]{WDS2016}%
  \BibitemOpen
  \bibfield  {author} {\bibinfo {author} {\bibfnamefont {F.}~\bibnamefont
  {Wong}}, \bibinfo {author} {\bibfnamefont {K.~K.}\ \bibnamefont {Dey}}, \
  and\ \bibinfo {author} {\bibfnamefont {A.}~\bibnamefont {Sen}},\ }\href
  {\doibase 10.1146/annurev-matsci-070115-032047} {\bibfield  {journal}
  {\bibinfo  {journal} {Annu. Rev. Mat. Res.}\ }\textbf {\bibinfo {volume}
  {46}},\ \bibinfo {pages} {407} (\bibinfo {year} {2016})}\BibitemShut
  {NoStop}%
\bibitem [{\citenamefont {Alarc\'on-Correa}\ \emph {et~al.}(2016)\citenamefont
  {Alarc\'on-Correa}, \citenamefont {Walker}, \citenamefont {Qiu},\ and\
  \citenamefont {Fischer}}]{AWQF16}%
  \BibitemOpen
  \bibfield  {author} {\bibinfo {author} {\bibfnamefont {M.}~\bibnamefont
  {Alarc\'on-Correa}}, \bibinfo {author} {\bibfnamefont {D.}~\bibnamefont
  {Walker}}, \bibinfo {author} {\bibfnamefont {T.}~\bibnamefont {Qiu}}, \ and\
  \bibinfo {author} {\bibfnamefont {P.}~\bibnamefont {Fischer}},\ }\href@noop
  {} {\bibfield  {journal} {\bibinfo  {journal} {Eur. Phys. J. Special Topics}\
  }\textbf {\bibinfo {volume} {225}},\ \bibinfo {pages} {2241} (\bibinfo {year}
  {2016})}\BibitemShut {NoStop}%
\bibitem [{\citenamefont {Gaspard}\ and\ \citenamefont {Kapral}(2019)}]{GK19}%
  \BibitemOpen
  \bibfield  {author} {\bibinfo {author} {\bibfnamefont {P.}~\bibnamefont
  {Gaspard}}\ and\ \bibinfo {author} {\bibfnamefont {R.}~\bibnamefont
  {Kapral}},\ }\href@noop {} {\bibfield  {journal} {\bibinfo  {journal} {Adv.
  Phys. X}\ }\textbf {\bibinfo {volume} {4}},\ \bibinfo {pages} {1602480}
  (\bibinfo {year} {2019})}\BibitemShut {NoStop}%
\bibitem [{\citenamefont {Gaspard}\ and\ \citenamefont
  {Kapral}(2018{\natexlab{a}})}]{GK18a}%
  \BibitemOpen
  \bibfield  {author} {\bibinfo {author} {\bibfnamefont {P.}~\bibnamefont
  {Gaspard}}\ and\ \bibinfo {author} {\bibfnamefont {R.}~\bibnamefont
  {Kapral}},\ }\href@noop {} {\bibfield  {journal} {\bibinfo  {journal} {J.
  Chem. Phys.}\ }\textbf {\bibinfo {volume} {148}},\ \bibinfo {pages} {134104}
  (\bibinfo {year} {2018}{\natexlab{a}})}\BibitemShut {NoStop}%
\bibitem [{\citenamefont {Lee}\ \emph {et~al.}(2014)\citenamefont {Lee},
  \citenamefont {Alarc{\'o}n-Correa}, \citenamefont {Miksch}, \citenamefont
  {Hahn}, \citenamefont {Gibbs},\ and\ \citenamefont {Fischer}}]{LAMHGF14}%
  \BibitemOpen
  \bibfield  {author} {\bibinfo {author} {\bibfnamefont {T.-C.}\ \bibnamefont
  {Lee}}, \bibinfo {author} {\bibfnamefont {M.}~\bibnamefont
  {Alarc{\'o}n-Correa}}, \bibinfo {author} {\bibfnamefont {C.}~\bibnamefont
  {Miksch}}, \bibinfo {author} {\bibfnamefont {K.}~\bibnamefont {Hahn}},
  \bibinfo {author} {\bibfnamefont {J.~G.}\ \bibnamefont {Gibbs}}, \ and\
  \bibinfo {author} {\bibfnamefont {P.}~\bibnamefont {Fischer}},\ }\href
  {\doibase 10.1021/nl500068n} {\bibfield  {journal} {\bibinfo  {journal} {Nano
  Lett.}\ }\textbf {\bibinfo {volume} {14}},\ \bibinfo {pages} {2407} (\bibinfo
  {year} {2014})}\BibitemShut {NoStop}%
\bibitem [{\citenamefont {Abdelmohsen}\ \emph {et~al.}(2014)\citenamefont
  {Abdelmohsen}, \citenamefont {Peng}, \citenamefont {Tu},\ and\ \citenamefont
  {Wilson}}]{APTW14}%
  \BibitemOpen
  \bibfield  {author} {\bibinfo {author} {\bibfnamefont {L.}~\bibnamefont
  {Abdelmohsen}}, \bibinfo {author} {\bibfnamefont {F.}~\bibnamefont {Peng}},
  \bibinfo {author} {\bibfnamefont {Y.}~\bibnamefont {Tu}}, \ and\ \bibinfo
  {author} {\bibfnamefont {D.~A.}\ \bibnamefont {Wilson}},\ }\href {\doibase
  10.1039/c3tb21451f} {\bibfield  {journal} {\bibinfo  {journal} {J. Mater.
  Chem. B}\ }\textbf {\bibinfo {volume} {2}},\ \bibinfo {pages} {2395}
  (\bibinfo {year} {2014})}\BibitemShut {NoStop}%
\bibitem [{\citenamefont {Colberg}\ and\ \citenamefont {Kapral}(2014)}]{CK14}%
  \BibitemOpen
  \bibfield  {author} {\bibinfo {author} {\bibfnamefont {P.}~\bibnamefont
  {Colberg}}\ and\ \bibinfo {author} {\bibfnamefont {R.}~\bibnamefont
  {Kapral}},\ }\href@noop {} {\bibfield  {journal} {\bibinfo  {journal}
  {Europhys. Lett.}\ }\textbf {\bibinfo {volume} {106}},\ \bibinfo {pages}
  {30004} (\bibinfo {year} {2014})}\BibitemShut {NoStop}%
\bibitem [{\citenamefont {Alder}\ and\ \citenamefont
  {Wainwright}(1967)}]{AW67}%
  \BibitemOpen
  \bibfield  {author} {\bibinfo {author} {\bibfnamefont {B.~J.}\ \bibnamefont
  {Alder}}\ and\ \bibinfo {author} {\bibfnamefont {T.~E.}\ \bibnamefont
  {Wainwright}},\ }\href {\doibase doi.org/10.1103/PhysRevLett.18.988}
  {\bibfield  {journal} {\bibinfo  {journal} {Phys. Rev. Lett.}\ }\textbf
  {\bibinfo {volume} {18}},\ \bibinfo {pages} {988} (\bibinfo {year}
  {1967})}\BibitemShut {NoStop}%
\bibitem [{\citenamefont {Alder}, \citenamefont {Gass},\ and\ \citenamefont
  {E.Wainwright}(1970)}]{AW70}%
  \BibitemOpen
  \bibfield  {author} {\bibinfo {author} {\bibfnamefont {B.~J.}\ \bibnamefont
  {Alder}}, \bibinfo {author} {\bibfnamefont {D.~M.}\ \bibnamefont {Gass}}, \
  and\ \bibinfo {author} {\bibfnamefont {T.}~\bibnamefont {E.Wainwright}},\
  }\href {\doibase doi.org/10.1103/PhysRevA.1.18} {\bibfield  {journal}
  {\bibinfo  {journal} {Phys. Rev. A}\ }\textbf {\bibinfo {volume} {1}},\
  \bibinfo {pages} {18} (\bibinfo {year} {1970})}\BibitemShut {NoStop}%
\bibitem [{\citenamefont {Dorfman}, \citenamefont {van Beijeren},\ and\
  \citenamefont {McClure}(1976)}]{DBM76}%
  \BibitemOpen
  \bibfield  {author} {\bibinfo {author} {\bibfnamefont {J.~R.}\ \bibnamefont
  {Dorfman}}, \bibinfo {author} {\bibfnamefont {H.}~\bibnamefont {van
  Beijeren}}, \ and\ \bibinfo {author} {\bibfnamefont {C.~F.}\ \bibnamefont
  {McClure}},\ }\href@noop {} {\bibfield  {journal} {\bibinfo  {journal} {Arch.
  Mech.}\ }\textbf {\bibinfo {volume} {28}},\ \bibinfo {pages} {333} (\bibinfo
  {year} {1976})}\BibitemShut {NoStop}%
\bibitem [{\citenamefont {Cukier}\ \emph {et~al.}(1980)\citenamefont {Cukier},
  \citenamefont {Kapral}, \citenamefont {Lebenhaft},\ and\ \citenamefont
  {Mehaffey}}]{CKLM80}%
  \BibitemOpen
  \bibfield  {author} {\bibinfo {author} {\bibfnamefont {R.~I.}\ \bibnamefont
  {Cukier}}, \bibinfo {author} {\bibfnamefont {R.}~\bibnamefont {Kapral}},
  \bibinfo {author} {\bibfnamefont {J.~R.}\ \bibnamefont {Lebenhaft}}, \ and\
  \bibinfo {author} {\bibfnamefont {J.~R.}\ \bibnamefont {Mehaffey}},\
  }\href@noop {} {\bibfield  {journal} {\bibinfo  {journal} {J. Chem. Phys.}\
  }\textbf {\bibinfo {volume} {73}},\ \bibinfo {pages} {5244} (\bibinfo {year}
  {1980})}\BibitemShut {NoStop}%
\bibitem [{\citenamefont {Schofield}\ and\ \citenamefont
  {Oppenheim}(1992)}]{SO92}%
  \BibitemOpen
  \bibfield  {author} {\bibinfo {author} {\bibfnamefont {J.}~\bibnamefont
  {Schofield}}\ and\ \bibinfo {author} {\bibfnamefont {I.}~\bibnamefont
  {Oppenheim}},\ }\href@noop {} {\bibfield  {journal} {\bibinfo  {journal}
  {Physica A}\ }\textbf {\bibinfo {volume} {18}},\ \bibinfo {pages} {187}
  (\bibinfo {year} {1992})}\BibitemShut {NoStop}%
\bibitem [{\citenamefont {Mazur}\ and\ \citenamefont {Oppenheim}(1970)}]{MO70}%
  \BibitemOpen
  \bibfield  {author} {\bibinfo {author} {\bibfnamefont {P.}~\bibnamefont
  {Mazur}}\ and\ \bibinfo {author} {\bibfnamefont {I.}~\bibnamefont
  {Oppenheim}},\ }\href@noop {} {\bibfield  {journal} {\bibinfo  {journal}
  {Physica}\ }\textbf {\bibinfo {volume} {50}},\ \bibinfo {pages} {241}
  (\bibinfo {year} {1970})}\BibitemShut {NoStop}%
\bibitem [{\citenamefont {Shea}\ and\ \citenamefont {Oppenheim}(1996)}]{SO96}%
  \BibitemOpen
  \bibfield  {author} {\bibinfo {author} {\bibfnamefont {J.-E.}\ \bibnamefont
  {Shea}}\ and\ \bibinfo {author} {\bibfnamefont {I.}~\bibnamefont
  {Oppenheim}},\ }\href@noop {} {\bibfield  {journal} {\bibinfo  {journal} {J.
  Phys. Chem}\ }\textbf {\bibinfo {volume} {100}},\ \bibinfo {pages} {19035}
  (\bibinfo {year} {1996})}\BibitemShut {NoStop}%
\bibitem [{\citenamefont {Espanol}\ and\ \citenamefont {Donev}(2015)}]{ED15}%
  \BibitemOpen
  \bibfield  {author} {\bibinfo {author} {\bibfnamefont {P.}~\bibnamefont
  {Espanol}}\ and\ \bibinfo {author} {\bibfnamefont {A.}~\bibnamefont
  {Donev}},\ }\href {\doibase dx.doi.org/10.1063/1.4936775} {\bibfield
  {journal} {\bibinfo  {journal} {J. Chem. Phys.}\ }\textbf {\bibinfo {volume}
  {143}},\ \bibinfo {pages} {234104} (\bibinfo {year} {2015})}\BibitemShut
  {NoStop}%
\bibitem [{col()}]{colloid}%
  \BibitemOpen
  \href@noop {} {}\bibinfo {note} {While a rigid colloidal particle is
  considered here, it is straightforward to replace it by a molecule or
  molecular aggregate with internal degrees of freedom.}\BibitemShut {Stop}%
\bibitem [{\citenamefont {van Zon}\ and\ \citenamefont
  {Schofield}(2007)}]{vanZS07}%
  \BibitemOpen
  \bibfield  {author} {\bibinfo {author} {\bibfnamefont {R.}~\bibnamefont {van
  Zon}}\ and\ \bibinfo {author} {\bibfnamefont {J.}~\bibnamefont {Schofield}},\
  }\href@noop {} {\bibfield  {journal} {\bibinfo  {journal} {J. Comp. Phys.}\
  }\textbf {\bibinfo {volume} {225}},\ \bibinfo {pages} {145} (\bibinfo {year}
  {2007})}\BibitemShut {NoStop}%
\bibitem [{\citenamefont {van Zon}\ and\ \citenamefont
  {Schofield}(2008)}]{vanZS08}%
  \BibitemOpen
  \bibfield  {author} {\bibinfo {author} {\bibfnamefont {R.}~\bibnamefont {van
  Zon}}\ and\ \bibinfo {author} {\bibfnamefont {J.}~\bibnamefont {Schofield}},\
  }\href {\doibase 10.1063/1.2901173} {\bibfield  {journal} {\bibinfo
  {journal} {J. Chem. Phys.}\ }\textbf {\bibinfo {volume} {128}},\ \bibinfo
  {pages} {154119} (\bibinfo {year} {2008})}\BibitemShut {NoStop}%
\bibitem [{\citenamefont {Goldstein}(1980)}]{Goldstein80}%
  \BibitemOpen
  \bibfield  {author} {\bibinfo {author} {\bibfnamefont {H.}~\bibnamefont
  {Goldstein}},\ }\href@noop {} {\emph {\bibinfo {title} {Classical
  Mechanics}}}\ (\bibinfo  {publisher} {Addison-Wesley},\ \bibinfo {address}
  {Reading, Massachusetts},\ \bibinfo {year} {1980})\BibitemShut {NoStop}%
\bibitem [{\citenamefont {Kapral}, \citenamefont {Consta},\ and\ \citenamefont
  {McWhirter}(1998)}]{KCM98}%
  \BibitemOpen
  \bibfield  {author} {\bibinfo {author} {\bibfnamefont {R.}~\bibnamefont
  {Kapral}}, \bibinfo {author} {\bibfnamefont {S.}~\bibnamefont {Consta}}, \
  and\ \bibinfo {author} {\bibfnamefont {L.}~\bibnamefont {McWhirter}},\ }in\
  \href@noop {} {\emph {\bibinfo {booktitle} {Classical and Quantum Dynamics in
  Condensed Phase Simulations}}},\ \bibinfo {editor} {edited by\ \bibinfo
  {editor} {\bibfnamefont {B.~J.}\ \bibnamefont {Berne}}, \bibinfo {editor}
  {\bibfnamefont {G.}~\bibnamefont {Ciccotti}}, \ and\ \bibinfo {editor}
  {\bibfnamefont {D.~F.}\ \bibnamefont {Coker}}}\ (\bibinfo  {publisher} {World
  Scientific},\ \bibinfo {address} {Singapore},\ \bibinfo {year} {1998})\ pp.\
  \bibinfo {pages} {583--616}\BibitemShut {NoStop}%
\bibitem [{\citenamefont {Carter}\ \emph {et~al.}(1989)\citenamefont {Carter},
  \citenamefont {Ciccotti}, \citenamefont {Hynes},\ and\ \citenamefont
  {Kapral}}]{CCHK89}%
  \BibitemOpen
  \bibfield  {author} {\bibinfo {author} {\bibfnamefont {E.}~\bibnamefont
  {Carter}}, \bibinfo {author} {\bibfnamefont {G.}~\bibnamefont {Ciccotti}},
  \bibinfo {author} {\bibfnamefont {J.~T.}\ \bibnamefont {Hynes}}, \ and\
  \bibinfo {author} {\bibfnamefont {R.}~\bibnamefont {Kapral}},\ }\href@noop {}
  {\bibfield  {journal} {\bibinfo  {journal} {Chem. Phys. Lett.}\ }\textbf
  {\bibinfo {volume} {156}},\ \bibinfo {pages} {472} (\bibinfo {year}
  {1989})}\BibitemShut {NoStop}%
\bibitem [{\citenamefont {Ciccotti}, \citenamefont {Kapral},\ and\
  \citenamefont {Vanden-Eijnden}(2005)}]{CKV05}%
  \BibitemOpen
  \bibfield  {author} {\bibinfo {author} {\bibfnamefont {G.}~\bibnamefont
  {Ciccotti}}, \bibinfo {author} {\bibfnamefont {R.}~\bibnamefont {Kapral}}, \
  and\ \bibinfo {author} {\bibfnamefont {E.}~\bibnamefont {Vanden-Eijnden}},\
  }\href {\doibase 10.1002/cphc.200400669} {\bibfield  {journal} {\bibinfo
  {journal} {ChemPhysChem}\ }\textbf {\bibinfo {volume} {6}},\ \bibinfo {pages}
  {1809} (\bibinfo {year} {2005})}\BibitemShut {NoStop}%
\bibitem [{\citenamefont {Robertson}(1967)}]{R67}%
  \BibitemOpen
  \bibfield  {author} {\bibinfo {author} {\bibfnamefont {B.}~\bibnamefont
  {Robertson}},\ }\href@noop {} {\bibfield  {journal} {\bibinfo  {journal}
  {Phys. Rev.}\ }\textbf {\bibinfo {volume} {160}},\ \bibinfo {pages} {175}
  (\bibinfo {year} {1967})}\BibitemShut {NoStop}%
\bibitem [{\citenamefont {Piccirelli}(1968)}]{P68}%
  \BibitemOpen
  \bibfield  {author} {\bibinfo {author} {\bibfnamefont {R.}~\bibnamefont
  {Piccirelli}},\ }\href@noop {} {\bibfield  {journal} {\bibinfo  {journal}
  {Phys. Rev.}\ }\textbf {\bibinfo {volume} {175}},\ \bibinfo {pages} {77}
  (\bibinfo {year} {1968})}\BibitemShut {NoStop}%
\bibitem [{\citenamefont {Oppenheim}\ and\ \citenamefont
  {Levine}(1979)}]{OL79}%
  \BibitemOpen
  \bibfield  {author} {\bibinfo {author} {\bibfnamefont {I.}~\bibnamefont
  {Oppenheim}}\ and\ \bibinfo {author} {\bibfnamefont {R.}~\bibnamefont
  {Levine}},\ }\href@noop {} {\bibfield  {journal} {\bibinfo  {journal}
  {Physica}\ }\textbf {\bibinfo {volume} {99A}},\ \bibinfo {pages} {383}
  (\bibinfo {year} {1979})}\BibitemShut {NoStop}%
\bibitem [{\citenamefont {Shea}\ and\ \citenamefont {Oppenheim}(1997)}]{SO97}%
  \BibitemOpen
  \bibfield  {author} {\bibinfo {author} {\bibfnamefont {J.-E.}\ \bibnamefont
  {Shea}}\ and\ \bibinfo {author} {\bibfnamefont {I.}~\bibnamefont
  {Oppenheim}},\ }\href@noop {} {\bibfield  {journal} {\bibinfo  {journal}
  {Physica A}\ }\textbf {\bibinfo {volume} {247}},\ \bibinfo {pages} {417}
  (\bibinfo {year} {1997})}\BibitemShut {NoStop}%
\bibitem [{\citenamefont {Shea}\ and\ \citenamefont {Oppenheim}(1998)}]{SO98}%
  \BibitemOpen
  \bibfield  {author} {\bibinfo {author} {\bibfnamefont {J.-E.}\ \bibnamefont
  {Shea}}\ and\ \bibinfo {author} {\bibfnamefont {I.}~\bibnamefont
  {Oppenheim}},\ }\href@noop {} {\bibfield  {journal} {\bibinfo  {journal}
  {Physica A}\ }\textbf {\bibinfo {volume} {250}},\ \bibinfo {pages} {265}
  (\bibinfo {year} {1998})}\BibitemShut {NoStop}%
\bibitem [{\citenamefont {Camargo}\ \emph {et~al.}(2018)\citenamefont
  {Camargo}, \citenamefont {de~la Torre}, \citenamefont {Duque-Zumajo},
  \citenamefont {Espanol}, \citenamefont {Delgado-Buscalioni},\ and\
  \citenamefont {Chejne}}]{CDDEDC18}%
  \BibitemOpen
  \bibfield  {author} {\bibinfo {author} {\bibfnamefont {D.}~\bibnamefont
  {Camargo}}, \bibinfo {author} {\bibfnamefont {J.~A.}\ \bibnamefont {de~la
  Torre}}, \bibinfo {author} {\bibfnamefont {D.}~\bibnamefont {Duque-Zumajo}},
  \bibinfo {author} {\bibfnamefont {P.}~\bibnamefont {Espanol}}, \bibinfo
  {author} {\bibfnamefont {R.}~\bibnamefont {Delgado-Buscalioni}}, \ and\
  \bibinfo {author} {\bibfnamefont {F.}~\bibnamefont {Chejne}},\ }\href
  {\doibase doi.org/10.1063/1.5010401} {\bibfield  {journal} {\bibinfo
  {journal} {J. Chem. Phys.}\ }\textbf {\bibinfo {volume} {148}},\ \bibinfo
  {pages} {064107} (\bibinfo {year} {2018})}\BibitemShut {NoStop}%
\bibitem [{\citenamefont {Camargo}\ \emph {et~al.}(2019)\citenamefont
  {Camargo}, \citenamefont {de~la Torre}, \citenamefont {Delgado-Buscalioni},
  \citenamefont {Chejne},\ and\ \citenamefont {Espanol}}]{CDDCE19}%
  \BibitemOpen
  \bibfield  {author} {\bibinfo {author} {\bibfnamefont {D.}~\bibnamefont
  {Camargo}}, \bibinfo {author} {\bibfnamefont {J.~A.}\ \bibnamefont {de~la
  Torre}}, \bibinfo {author} {\bibfnamefont {R.}~\bibnamefont
  {Delgado-Buscalioni}}, \bibinfo {author} {\bibfnamefont {F.}~\bibnamefont
  {Chejne}}, \ and\ \bibinfo {author} {\bibfnamefont {P.}~\bibnamefont
  {Espanol}},\ }\href {\doibase 10.1063/1.5088354} {\bibfield  {journal}
  {\bibinfo  {journal} {J. Chem. Phys.}\ }\textbf {\bibinfo {volume} {150}},\
  \bibinfo {pages} {144104} (\bibinfo {year} {2019})}\BibitemShut {NoStop}%
\bibitem [{\citenamefont {Kavassalis}\ and\ \citenamefont
  {Oppenheim}(1988)}]{KO88}%
  \BibitemOpen
  \bibfield  {author} {\bibinfo {author} {\bibfnamefont {T.~A.}\ \bibnamefont
  {Kavassalis}}\ and\ \bibinfo {author} {\bibfnamefont {I.}~\bibnamefont
  {Oppenheim}},\ }\href@noop {} {\bibfield  {journal} {\bibinfo  {journal}
  {Physica}\ }\textbf {\bibinfo {volume} {148A}},\ \bibinfo {pages} {521}
  (\bibinfo {year} {1988})}\BibitemShut {NoStop}%
\bibitem [{\citenamefont {Schofield}, \citenamefont {Lim},\ and\ \citenamefont
  {Oppenheim}(1992)}]{SLO92}%
  \BibitemOpen
  \bibfield  {author} {\bibinfo {author} {\bibfnamefont {J.}~\bibnamefont
  {Schofield}}, \bibinfo {author} {\bibfnamefont {R.}~\bibnamefont {Lim}}, \
  and\ \bibinfo {author} {\bibfnamefont {I.}~\bibnamefont {Oppenheim}},\ }\href
  {\doibase 10.1016/0378-4371(92)90198-y} {\bibfield  {journal} {\bibinfo
  {journal} {Physica A}\ }\textbf {\bibinfo {volume} {181}},\ \bibinfo {pages}
  {89} (\bibinfo {year} {1992})}\BibitemShut {NoStop}%
\bibitem [{\citenamefont {Schofield}\ and\ \citenamefont
  {Oppenheim}(1994)}]{SO94}%
  \BibitemOpen
  \bibfield  {author} {\bibinfo {author} {\bibfnamefont {J.}~\bibnamefont
  {Schofield}}\ and\ \bibinfo {author} {\bibfnamefont {I.}~\bibnamefont
  {Oppenheim}},\ }\href@noop {} {\bibfield  {journal} {\bibinfo  {journal}
  {Physica A}\ }\textbf {\bibinfo {volume} {204}},\ \bibinfo {pages} {555}
  (\bibinfo {year} {1994})}\BibitemShut {NoStop}%
\bibitem [{mas()}]{mass-scaling}%
  \BibitemOpen
  \href@noop {} {}\bibinfo {note} {Although the domain of validity of these
  equations is more restricted than that of the generalized Langenin
  equation~(37) because of the assumption $M \gg m$, even for colloids with
  linear dimensions of a few nanometers there can be a large mass dispartity
  between the colloid and solvent particles.}\BibitemShut {Stop}%
\bibitem [{\citenamefont {Gaspard}\ and\ \citenamefont
  {Kapral}(2018{\natexlab{b}})}]{GK18b}%
  \BibitemOpen
  \bibfield  {author} {\bibinfo {author} {\bibfnamefont {P.}~\bibnamefont
  {Gaspard}}\ and\ \bibinfo {author} {\bibfnamefont {R.}~\bibnamefont
  {Kapral}},\ }\href@noop {} {\bibfield  {journal} {\bibinfo  {journal} {J.
  Chem. Phys.}\ }\textbf {\bibinfo {volume} {148}},\ \bibinfo {pages} {194114}
  (\bibinfo {year} {2018}{\natexlab{b}})}\BibitemShut {NoStop}%
\bibitem [{\citenamefont {Mazo}(1967)}]{M67}%
  \BibitemOpen
  \bibfield  {author} {\bibinfo {author} {\bibfnamefont {R.~M.}\ \bibnamefont
  {Mazo}},\ }\href@noop {} {\emph {\bibinfo {title} {Statistical Mechanical
  Theories of Transport Processes}}}\ (\bibinfo  {publisher} {Pergamon Press},\
  \bibinfo {address} {New York},\ \bibinfo {year} {1967})\BibitemShut {NoStop}%
\end{thebibliography}%

\end{document}